\documentclass[12pt,english]{article}
\PassOptionsToPackage{natbib=true}{biblatex}
\usepackage[T1]{fontenc}
\usepackage[utf8]{inputenc}
\usepackage{booktabs}
\usepackage{url}
\usepackage{multirow}
\usepackage{amsmath}
\usepackage{amsthm}
\usepackage{graphicx}
\usepackage{setspace}
\makeatletter

\usepackage{chenpaper}

\usepackage{babel}

\makeatother

\usepackage{babel}
\usepackage[style=authoryear,maxcitenames=3, mincitenames=1,uniquename=false]{biblatex}
\begin{document}
\title{\textbf{High-Throughput Asset Pricing}}
\author{{Andrew Y. Chen}\\
{\normalsize Federal Reserve Board}\and {Chukwuma Dim}\\
{\normalsize George Washington University}}
\date{June 2025\thanks{Alternative title: ``How I learned to stop worrying and love data
mining.'' First posted to arXiv.org: November 2023. Replication code:
\protect\protect\url{https://github.com/chenandrewy/high-throughput-ap}.
Long-short strategy data: \protect\protect\url{https://sites.google.com/site/chenandrewy/}. We thank our discussants: Rudiger Weber, Julio Crego, and Morad Zekhnini for helpful comments. We also thank participants at the TBEAR Network Workshop (KIT), FutFinInfo, MFA, Iowa State, and Baylor. The views expressed herein are those of the authors and do not necessarily
reflect the position of the Board of Governors of the Federal Reserve
or the Federal Reserve System.}}

\maketitle
 
\begin{abstract}
\begin{singlespace}
\noindent We apply empirical Bayes (EB) to mine data on 136,000 long-short strategies constructed from accounting ratios, past returns, and ticker symbols. This ``high-throughput asset pricing'' matches the out-of-sample performance of top journals while eliminating look-ahead bias. Naively mining for the largest Sharpe ratios leads to similar performance, consistent with our theoretical results, though EB uniquely provides unbiased predictions with transparent intuition. Predictability is concentrated in accounting strategies, small stocks, and pre-2004 periods, consistent with limited attention theories. Multiple testing methods popular in finance fail to identify most out-of-sample performers. High-throughput methods provide a rigorous, unbiased framework for understanding asset prices.
\end{singlespace}
\end{abstract}
\vspace{10ex}
 \textbf{JEL Classification}: G0, G1, C1

\noindent\textbf{Keywords}: stock market predictability,
stock market anomalies, p-hacking, multiple testing \thispagestyle{empty}\setcounter{page}{0}

\vspace{10ex}

\pagebreak

\section{Introduction}

\setcounter{page}{1}

Data mining refers to searching data for interesting patterns. This
search leads to data mining bias, if many patterns are just chance
results, as is surely the case with stock return data. To address
this problem, the asset pricing literature recommends restricting
the search to patterns consistent with theory (\citet*{cochrane2005risk,harvey2017presidential}).
However, recent empirical evidence finds this method is ineffective,
even for theories published in top finance journals (\citet*{chen2022peer}).

We offer a different solution. Instead of mining data less, we recommend
mining data \emph{rigorously}. Rigorous data mining means conditioning
interesting results on the fact that they come from searching through
data. This conditioning can be achieved using empirical Bayes (\citet{robbins1956empirical,efron1973stein,efron2012large}).
Rigorous data mining also means that the search should be systematic,
as is commonly done in high-throughput biology and chemistry (\citet{yang2021high}).
Ironically, systematic search implies that asset pricing should involve
\emph{more} data mining, not less.

We use empirical Bayes (EB) to mine for out-of-sample returns among
136,000 long-short trading strategies. The trading strategies are
constructed from systematically searching data on accounting ratios,
past returns, and stock tickers. Through this ``high-throughput asset
pricing,'' we construct a portfolio with out-of-sample returns that
are comparable to the returns from the best journals in finance.

Our data-mined portfolio is the simple average of the top 1\% of strategies,
based on EB-predicted Sharpe ratios. It earns out-of-sample returns
of 5.7\% per year over the 1983-2020 sample, compared to the mean
return of 5.9\% per year found by averaging the 200 published strategies from \citet{ChenZimmermann2021}. But unlike the published strategies, which
were selected with knowledge of stock return patterns that occurred
in the 1980s and 1990s, our strategies can be constructed using only
information available in real time.  

In fact, even naively mining for the largest Sharpe ratios leads to publication-like performance. We provide a theoretical explanation for this phenomenon in Proposition \ref{prop:optimal-naive}, which shows that under  standard statistical practices (\citealt{fisher1925statistical}), naive data mining often selects the same set of strategies as an ideal Bayesian. However, while the naively-selected strategies may be optimal, naive performance estimates are distorted, illustrating the importance of rigorously data mining with EB.

The top 1\% portfolio selected by EB provides insights into the
nature of return predictability. 91.0\% of strategies in this portfolio
are equal-weighted accounting ratio strategies. Almost all of the
remainder are equal-weighted past-return strategies. Moreover, the
returns of the top 1\% strategy are concentrated in the pre-2004 data.
These facts are consistent with the theory that predictability is
largely due to limited attention and the slow incorporation of information
into stock prices (\citet{peng2005learning}; \citet{Chordia2014Have}).

Other facts shed light on the drivers of the recent decline in cross-sectional
predictability. We find that the returns of the top 5\% and top 10\%
of portfolios are also concentrated in the pre-2004 data. These strategies
are enormous in number: the top 5\% consists of 6,305 strategies,
and the top 10\% consists of 12,610. As many of these strategies are
unlikely to be found in academic journals, this suggests that the
key driver of the recent declines in predictability is improvements
in information technology (\citet{Chordia2014Have}), rather than
investors learning from academic publications (\citet{Mclean2016Does}).
Consistent with this idea, we find that the top 20 strategies according
to predicted Sharpe ratios using data available in 1993 have themes
rarely seen in academic journals, like mortgage debt, growth in interest
expense, and depreciation. Themes that were popular in academia in
1993, like book-to-market, momentum, and sales growth are missing
from this list.

Overall, high-throughput asset pricing provides not only a method
for dealing with look-ahead bias, but also a more rigorous method
for documenting asset pricing facts. We post our strategy returns
and code publicly, and encourage future researchers to use these methods.

Unlike many big data methods, EB provides a transparent
intuition. In essence, EB measures the distance between
the empirical t-stat distribution and the standard normal null. Ticker-based
strategies have t-stats that are extremely close to the null, implying
no predictability. In contrast, equal-weighted accounting t-stats
are too fat tailed to be consistent with the null, implying strong
predictability. Thus, just by visually inspecting the t-stat distributions,
one can see where predictability is concentrated.

EB provides highly accurate predictions in pre-2004 data. We construct
120 portfolio tests using the 136,000 data-mined strategies, and compare
EB-predicted returns with out-of-sample returns. In almost all of
the 120 portfolios, the EB predictions are within 2 standard errors
of the out-of-sample mean.

Post-2004, EB has more difficulty with accuracy, though it still captures
broad patterns in out-of-sample returns. Compared to pre-2004, predicted
returns are closer to zero, and only equal-weighted accounting strategies
show notable predicted returns. However, out-of-sample returns are
even closer to zero than predicted. This difficulty might be expected
given the rise of information technology around 2004, which likely
led to a structural break in predictability (\citealt{Chordia2014Have}; \citealt{kim2021causal}).
Our EB predictions are constructed using a simple 20-year rolling
window, and thus fail to account for this break. This difficulty suggests
that a smart data miner armed with theory might have understood the
implications of the internet, and could perhaps have performed much
better than our theory-free EB mining process.

We also illustrate how improper use of multiple testing statistics can lead to poor data mining results. We demonstrate this possibility using \citepos{harvey2016and} recommended method for false discovery control. \citet{harvey2016and} recommend applying \citepos{benjamini2001control} Theorem 1.3 to construct a t-stat hurdle that controls the false discovery rate (FDR) at the 1\% level. Nearly all of our 136,000 trading strategies fail to meet this hurdle, suggesting that there are few interesting patterns in this data. But in fact, simple out-of-sample tests show there are thousands of strategies with notable out-of-sample returns. We find similar results following the recommended multiple testing control in \citet{chordia2020anomalies}, which is based on \citet{romano2007control}. In contrast, the \citet{storey2002direct} FDR control recommended in \citet{barras2010false}, captures the majority of notable portfolios. 

Fortunately, this error can be avoided by rigorously studying the statistics. According to \citet{benjamini2001control}, their Theorem 1.3 is ``very often unneeded, and yields too conservative of a procedure.'' This negative sentiment is echoed in Efron's \citeyearpar{efron2012large} textbook on large scale inference. In contrast, the EB methods we use are recommended for settings like ours in Chapter 1 of \citet{efron2012large}, as well as Chapters 6 and 7 of \citet{efron2016computer}.\footnote{A brief explanation of why \citet{benjamini2001control} Theorem 1.3 is excessively conservative is found in Section 2.5 of \citet{chen2024t}.} The statistics literature has relatively little to say about the method recommended in \citet{chordia2020anomalies}. We provide our own characterization, which illustrates how this method is appropriate if selecting a null strategy is catastrophic.  But using the standard null, that the mean long-short return (or alpha) is zero, Chordia et al.'s method implies unneeded conservatism.

\subsection{Related Literature}

We add to \citet{yan2017fundamental} and \citet{chen2022peer}, who
document that mining accounting data can produce substantial out-of-sample
returns. Accounting data is important: Chen et al. find that mining
ticker variables leads to out of sample returns of approximately zero.
Thus, one needs a method for identifying the predictive power of accounting
data in real time. Our empirical Bayes formulas provide one such method.

The literature on multiple testing in asset pricing features disagreement
on both the methods that should be used and the empirical extent of
multiple testing problems. \citet{chen2020publication}; \citet{chen2022zeroing};
and \citet{jensen2023there} recommend empirical Bayes shrinkage.
In contrast, \citet{harvey2016and}; \citet{harvey2020false}; and
\citet{chordia2020anomalies} recommend conservative false discovery
controls, much more conservative than the FDR methods in \citet{barras2010false}.
We show how empirical Bayes shrinkage and the recommended method from
\citet{barras2010false} leads to much more accurate inferences. More recently, \citet{marrow2024real} use empirical Bayes to study past return signals, with a focus on signal interactions and optimal weighting of more recent data.

In contrast to the intuition that simplicity is a virtue, we find
that studying an enormous number of potential predictors leads to
insights about the nature of return predictability. A similar theme
is found in \citet{kelly2024virtue} and \citet{didisheim2023complexity},
who illustrate the ``virtue of complexity'' in the modeling of expected
returns.

\section{Data and Methods\protect}\label{sec:methods}

We describe the data (Section \ref{sec:data}) and how we rigorously
mine it (Sections \ref{sec:stats:eb-overview}-\ref{sec:stats:eb-details}).

\subsection{Data on 136,000 Trading Strategies\protect}\label{sec:data}

Table \ref{tab:data-descrip} describes our data-mined strategies.
The strategies are either based on accounting ratios, past returns,
or tickers. Accounting ratio strategies are taken from \citet{chen2022peer}.\footnote{We are grateful that the authors make their data publicly available.}
The past return and ticker strategies are inspired by \citet{yan2017fundamental}
and \citet{harvey2017presidential}, respectively, but we generate
our own strategies in order to ensure that the number of strategies
is comparable across data sources and to ensure that each type of
strategy consists of many distinct strategies.\footnote{Results that mine data following \citet{yan2017fundamental} and \citet{harvey2017presidential}
are similar and can be found in the first draft of our paper on arxiv.org
or via our github site.} 
\begin{center}
{[}Table \ref{tab:data-descrip}, Overview of Trading Strategies,
about here{]} 
\par\end{center}

A key feature of these strategies is that they are \emph{not }selected
based on having notable historical returns. Instead, they are constructed
to systematically explore various types of data. So unlike most datasets
in asset pricing (e.g. Ken French's size- and B/M-sorted portfolios;
\citet{ChenZimmermann2021}), ours is arguably free of data mining
bias. Indeed, Table \ref{tab:data-descrip} shows that the median
sample mean return is close to zero for all sets of strategies.

In high-throughput research, the median measurement is relatively
unimportant. What matters is that the extreme measurements show promise
for, say, a pharmaceutical intervention or cancer prediction. The
extreme measurements in Table \ref{tab:data-descrip} suggest that
accounting and past return data show promise for predicting returns.
These data lead to mean returns that can exceed 5 percent per year
in absolute value.

For further details on the strategy definitions, see Appendix \ref{sec:app:data}
or our github site.

\subsection{Empirical Bayes Overview}\label{sec:stats:eb-overview}

The 136,000 strategies in Table \ref{tab:data-descrip} contain the
potential for significant data mining bias. To understand the bias, let $r_{i}$ be a performance measure for strategy $i$ (e.g. mean return, alpha) and decompose it as follows: 
\begin{align}\label{eq:r=mu+e}
r_{i} & =\mu_{i}+\varepsilon_{i}
\end{align}
where $\mu_{i}$ is the actual performance and $\varepsilon_{i}$ is sampling error or luck.

Data mining involves selecting $i$ with large $r_{i}$. Suppose we set $\bar{r} \gg 0$, and  search for $i^\ast \in \{1;2;\ldots;136,000\}$ such that $r_{i^\ast} = \bar{r}$. This practice is dangerous because one might think $\bar{r}$ is a good estimate of $\mu_{i^\ast}$. However, $\bar{r}$ is in fact biased upward
\begin{align} \label{eq:bias-demo}
\bar{r} &= E\left(r_{i^\ast} | r_{i^\ast} = \bar{r}\right) \\
 &= E\left(\mu_{i^\ast} | r_{i^\ast} = \bar{r}\right) 
 + \underbrace{
    E\left(\varepsilon_{i^\ast} | r_{i^\ast} = \bar{r}\right)
    }_{>0}  
    > \mu_{i^\ast}. \notag
\end{align}
Selecting for large $r_{i}$ also selects for large $\varepsilon_{i}$, leading to $E\left(\varepsilon_{i^\ast} | r_{i^\ast} = \bar{r}\right) > 0$ and the bias in Equation (\ref{eq:bias-demo}).

To data mine safely, one needs to remove the luck term $E\left(\varepsilon_{i}|r_{i^\ast} = \bar{r}\right)$. This term is just a conditional expectation, so it can be computed using Bayes rule, provided one has a probability model for $\mu_{i}$ and $r_{i}$.

Suppose one has a probability model, with parameter vector $\Omega$. The bias can then be removed by computing
\begin{align}\label{eq:luck-sketch}
    E\left(\mu_{i^\ast}|r_{i^\ast} = \bar{r};\hat{\Omega}\right)
    = \bar{r} - E\left(\varepsilon_{i^\ast}|r_{i^\ast} = \bar{r};\hat{\Omega}\right)
\end{align}
where $\hat{\Omega}$ is a consistent (frequentist) estimate of the
probability model parameters. This method, of applying frequentist  estimates to Bayesian formulas is known as ``empirical Bayes'' (\citet{robbins1956empirical,efron1973stein}).

Equation (\ref{eq:luck-sketch}) conditions on only one statistic regarding strategy $i$. A more optimal estimate uses more information
\begin{align}\label{eq:ephat-def}
    E\left(
        \mu_{i^\ast}|r_{i^\ast} = \bar{r}, X_{i} = \bar{X};\hat{\Omega}
    \right)
    = 
    \bar{r} 
    - E\left(
        \varepsilon_{i^\ast}|r_{i^\ast} = \bar{r}, X_{i} = \bar{X};\hat{\Omega}
    \right)
\end{align}
where $X_{i}$ is a vector of additional statistics for strategy $i$ and $\bar{X}$ is a realized value of $X_{i}$. For example, $X_{i}$ can include the standard error of $r_{i}$, the portfolio weighting (equal- or value-weighted), and the signal data source (accounting, past returns, tickers).

We use Equation (\ref{eq:ephat-def}) to search our 136,000 strategies for large expected returns. We will not use economic theory
to determine the probability model, and thus our search is largely
atheoretical. However, we recognize the bias that comes from such
a search (Equation (\ref{eq:bias-demo})), and carefully correct for
it. Thus, we describe our methods as ``rigorous data mining.''

\subsection{Optimal Naive Data Mining}\label{sec:stats:naive-theory}

In empirical asset pricing, we are often interested in two questions:
\begin{enumerate}
\item What are the best strategies? 
\item What is the performance of the best strategies? 
\end{enumerate}
If one is interested \emph{only} in the first question, then there is a sense in which naively mining data, without accounting for data mining bias, is often optimal. 

To understand this, we add structure to the model. First, explicitly define the additional statistics $X_i$:
\begin{align}
    X_i &= \left[ D_i, \SE_i \right]
\end{align}
where $D_i$ is the strategy ``family'' (e.g. equal-weighted accounting) and $\SE_i$ is the standard error of $r_i$. Actual performance follows
\begin{align}\label{eq:mu-X}
    \mu_{i}|X_{i}	&\sim g_{D_{i}, \SE_{i}}\left(\cdot\right)
\end{align}
where $g_{D_{i}, \SE_{i}}\left(\cdot\right)$ is a distribution that depends on $D_i$ and $\SE_i$. Measured performance follows
\begin{align}
    r_{i}|\mu_{i}, X_{i}	&\sim f_{\mu_{i},\SE_{i}}\left(\cdot\right)
\end{align}
where $f_{\mu_{i},\SE_{i}}\left(\cdot\right)$ is a distribution that depends on $\mu_{i}$ and $\SE_{i}$. This is a hierarchical structure, where the strategy family determines the actual performance, which in turn determines the measured performance.  

Second, define data mining. Naive data mining chooses a hurdle $h$ and then selects strategies
\begin{align}\label{eq:naive-selection}
    \{i: r_{i} > h\}
\end{align}
In contrast, EB data mining uses the bias-adjusted measure to select strategies
\begin{align}\label{eq:eb-selection}
    \{i: E\left(\mu_{i}\mid r_{i},X_{i}\right) > h^\prime\}
\end{align}
where $h^\prime$ is chosen to select the same number of strategies as in naive data mining.

In general, Equations (\ref{eq:naive-selection}) and (\ref{eq:eb-selection}) imply different sets of strategies. However, under some natural conditions, the selections are identical:

\begin{proposition}\label{prop:optimal-naive}
    Consider the following two conditions:
    \begin{enumerate}
        \item \label{cond:normal} The performance measure satisfies
        \begin{align}
            r_{i} \mid \mu_{i}, X_{i} 
            \sim \text{Normal}\left(\mu_{i}, \SE^2\right)
        \end{align}
        where $\SE$ is a constant.
        \item \label{cond:tail} The hurdle $h$ satisfies
        \begin{align}
            \Pr\left(r_{i}>h|D_{i}\right)&=0
            \quad\text{if }D_{i}\in\mathcal{D} \label{eq:tail-cond1} \\
            \mu_{i}|X_{i}	&\sim g_{\SE_{i}}\left(\cdot\right) 
            \quad\text{if }D_{i}\in\mathcal{D} \label{eq:tail-cond2}    
        \end{align}
        where $\mathcal{D}$ is a subset of the possible strategy families and $g_{\SE_{i}}\left(\cdot\right)$ is distribution with positive variance that does not depend on $D_{i}$. 
    \end{enumerate}
    If conditions \ref{cond:normal} and \ref{cond:tail} hold, then naive data mining selects the same set of strategies as empirical Bayes.
\end{proposition}

Conditions \ref{cond:normal} and \ref{cond:tail} arise naturally when using long samples (e.g. 300 months of returns), standardized performance measures (e.g. t-statistics), and strict statistical hurdles (e.g. 5\% critical levels). Under these conditions, actual performance is a strictly increasing function of only the measured performance, as proved in Appendix \ref{sec:app:proof}. As a result, data-mined performance provides a reliable signal of actual performance, even if the magnitudes are distorted. The proposition assumes some exact conditions, and leading to identical selections, but approximate conditions would likely lead to similar selections. 

One interpretation of Proposition \ref{prop:optimal-naive} is that Fisher's \citeyearpar{fisher1925statistical} focus on t-statistics set future researchers up for success, even in the modern era of big data. 

On the other hand, Fisher would likely have been unsatisfied with finding the best strategies. He most likely would implore us to find unbiased estimates for these best performers. Thus to rigorously mine data, one should still apply empirical Bayes.

\subsection{Empirical Bayes Implementation \protect}\label{sec:stats:eb-details}

We select as our performance measure the t-statistic on the raw long-short return, and assume that standard errors are precisely measured, implying
\begin{align}
    r_{i}|\mu_{i}, X_{i} &\sim \text{Normal}\left(\mu_{i}, 1\right)
    \label{eq:t=theta+delta}
\end{align}
The latent performance is a mixture of two normals that depends on the strategy family $D_i$.
\begin{align}
\mu_{i}\mid (X_i , D_i = d)
& \sim\begin{cases}
\text{Normal}\left(
    \theta_{d,1}, \sigma_{d,1}^2
\right) 
    & \text{with prob }\lambda_{d}\\
\text{Normal}\left(
    \theta_{d,2},\sigma_{d,2}^{2}
\right) 
    & \text{otherwise}
\end{cases}.
\label{eq:theta~mixnorm}
\end{align}
where $d$ is one of the six strategy families that comes from combining three data sources (accounting, past returns, tickers) with two portfolio formation methods (equal-weighted and value-weighted). Mixture normals are parsimonious, easy to understand, and yet allow for skewness and fat tails. 

We then estimate $\Omega\equiv\left[\theta_{d,1}, \sigma_{d,1}^2, \theta_{d,2}, \sigma_{d,2}^2, \lambda_d\right]_{d=1,\ldots,6}$ using quasi-maximum likelihood. The quasi-likelihood is computed using the distr package (\citet{ruckdeschel2006s4}). Optimization of $\Omega$ uses nloptr (\citet{NLopt}).This estimation is done using the past 20 years of long-short returns, separately for each ``forecasting year''  spanning 1983-2019.

Finally, we recover EB predictions by computing Equation (\ref{eq:ephat-def}) with distr, which produces an EB prediction of the expected return in units of standard errors. The EB predicted return is just Equation (\ref{eq:ephat-def}) multiplied by the standard error. Similarly, the EB predicted Sharpe ratio is Equation (\ref{eq:ephat-def}) multiplied by the square root of the number of periods in the sample.

For further details see Appendix
\ref{sec:app:theory} or our github site.

\section{Performance of the Best Data Mined Strategies\protect}\label{sec:best}

We show that data mining leads to research-like out-of-sample
returns (Section \ref{sec:best-oos}) and take a look at which kinds
of strategies are identified by data mining (Section \ref{sec:best-detail}).
We also provide intuition for why data mining produces such high returns (Section \ref{sec:best-intuition}).

\subsection{Out-of-Sample Returns\protect}\label{sec:best-oos}

Can data mining generate out-of-sample returns? To answer this question we construct simple out-of-sample portfolio tests.

Each year, we sign strategies to have positive predicted returns,
and then form portfolios that equally-weight strategies in the top
$X\%$ of predicted Sharpe ratios. We use both EB predictions and standard naive predictions.  We examine $X=$  1, 5, and 10.
For comparison, we also examine a portfolio that equally-weighs published
strategies from the \citet{ChenZimmermann2021} dataset.

Table \ref{tab:beststrats} shows the result. Using empirical Bayes (EB Mining), the top 1\% of strategies perform similarly to strategies published in top finance
journals. Over the full 1983-2020 sample, the top 1\% portfolio earns
5.70\% per year, compared to the 5.88\% return from published strategies.
The Sharpe ratio from EB mining is smaller, at 1.46 vs 2.03 for
published strategies. However, unlike the EB-mined strategies, which
are formed using only information available in real-time, the published
strategies contain look-ahead bias. Indeed, if we focus on strategies
in top journals that were published pre-2004, the performance is very
similar to the EB-mined strategies in terms of either mean returns
or Sharpe ratios. 
\begin{center}
{[}Table \ref{tab:beststrats}, Returns of Long-Short Portfolios Data-Mined,
about here{]} 
\par\end{center}

The EB-mined returns are robust. The top 5\% and top 10\% of data-mined
strategies also perform well and are extremely statistically significant, indicating that the performance of the top 1\% is not driven by outliers. 

Panel B shows that even naive data mining produces research-like returns. Simply choosing the top 1\% of strategies based on their past Sharpe ratios leads to an out-of-sample Sharpe ratio of 1.45. This is almost exactly the same as the Sharpe ratio from the top 1\% using EB mining, consistent with Proposition \ref{prop:optimal-naive}. The top 5\% and top 10\% of naive strategies underperform a bit relative to EB mining, but the intuition behind Proposition \ref{prop:optimal-naive} still goes through.

Figure \ref{fig:beststrats-cret} takes a closer look by plotting
the value of \$1 invested in each portfolio over time. The top 1\%
data-mined strategies have similar performance to published strategies
throughout the figure. All portfolios show relatively little
cyclicality during the recessions of 1991, 2009, and 2020. Indeed,
the returns are fairly consistent throughout the chart, with an important
caveat. 
\begin{center}
{[}Figure \ref{fig:beststrats-cret}, Cumulative Long-Short Returns,
about here{]} 
\par\end{center}

The caveat is that returns are concentrated in the pre-2004 sample.
This is seen in the flattening of the solid line in Figure \ref{fig:beststrats-cret} around 2004.  The EB Mining top 1\% portfolio returns 8.17\% per year from 1983-2005, compared to just 2.03\% from 2005-2020. A similar decay is seen across all portfolios, both data-mined and academic. This decay is consistent with \citet{Chordia2014Have} and \citet{chen2022zeroing}, who argue that the rise of information technology reduced return predictability.

Overall, we find that one can find long-short returns comparable to
those from the best journals in finance, just by mining data, with
little thought about the underlying economics. Moreover, rigorous
data mining can discriminate between data sources that have no information
about future returns, like stock market tickers, from data that is
rich in information, like accounting ratios. Unlike the published
strategy returns, our returns can be found using only information
available in real-time. These results show that high-throughput methods
provide a bias-free approach to studying stock market predictability.
Our strategy returns and code are public, and we encourage future
researchers to use these methods.

\subsection{The Composition of the Top 1\%\protect}\label{sec:best-detail}

Table \ref{tab:top_strat_desc} takes a closer look at the top 1\%
strategies produced by rigorous data mining. Panel A shows that 91.0\%
of the top 1\% come from the equal-weighted accounting family and
8.6\% come from equal-weighted past returns. The other strategy families
comprise a negligible part of the top 1\%. Ticker strategies are completely
absent. 
\begin{center}
{[}Table \ref{tab:top_strat_desc}, Description of Top 1\% Data-Mined
Strategies, about here{]} 
\par\end{center}

Taken with Table \ref{tab:beststrats}, these results show that cross-sectional
predictability is concentrated in accounting data, small stocks, and
pre-2004 samples. These stylized facts offer a parsimonious description
of the ``factor zoo.'' Theories that wish to capture the big picture
of cross-sectional predictability should be consistent with these facts.
For example, slow diffusion of economic information is consistent,
as this diffusion would be especially slow in small stocks and before
the internet era. In this way, high throughput asset pricing provides
a way to not only identify out-of-sample returns, but to also provide
insight into the underlying economics.

Panel B of Table \ref{tab:top_strat_desc} shows that many of the top 1\% strategies are quite far from
the predictors noted in the academic literature. In 1993, academics
were focused on predictors like book-to-market, 12-month momentum,
and sales growth (\citet{fama1992cross}; \citet{jegadeesh1993returns};
\citet{lakonishok1994contrarian}). None of these predictors are in
the top 20 strategies based on predicted Sharpe ratios from rigorous
data mining. Instead, the common themes from data mining include shorting
stocks with high or growing debt, as well as buying stocks with high
depreciation, depletion, and amortization. Another theme is buying
stocks with high returns in quarters $t$ minus 17 and 18. 

Based on textbook risk-based or behavioral asset pricing, one might
expect that these data-mined predictors will average zero returns
out-of-sample. But this is not the case. The realized Sharpe ratios
for these strategies in the 10 years after 1993 averages around 1.0
(``SR OOS'' column). 

Panel B of Table \ref{tab:top_strat_desc} focuses on 1993 because well-known predictability papers were published around that time (e.g. \citet{fama1993common}). In other years, the top 20 list is different, though shorting variables related to debt growth remains a common theme. For further details see 
Appendix Tables \ref{tab:top_strat_alt1} and \ref{tab:top_strat_alt2}.

\subsection{Shrinkage Intuition\protect}\label{sec:best-intuition}

Unlike many big data and machine learning methods, empirical Bayes
has a transparent intuition. The intuition can be seen in a special
case of the prediction Equation (\ref{eq:ephat-def}). If $\mu_{i}\mid \left(X_i, D_i = d\right) \sim\text{Normal}\left(0,\sigma_{d}^{2}\right)$,
we have 
\begin{align}
E\left(\mu_i \mid r_i = \bar{r}, X_i = \bar{X}, D_i = d\right)
& =\left[
    1-\frac{1}{\widehat{\Var}\left(r_{i} | D_i = d\right)}
\right]
\bar{r},\label{eq:shrink-ez}
\end{align}
where $\widehat{\Var}\left(r_{i} | D_i = d\right)$ is an estimate of the cross-strategy variance of performance measures among strategies with data family $d$. 

This expression says that rigorous mining involves shrinking performance measures $r_{i}$ toward zero at a rate of $\frac{1}{\widehat{\Var}\left(r_{i} | D_i = d\right)}$. $\widehat{\Var}\left(r_{i} | D_i = d\right)$ measures how far the data are from the null of $r_{i} \sim\text{Normal}\left(0,1\right)$, which we imposed in Equation (\ref{eq:t=theta+delta}). If there is no predictability, then $r_{i}\sim\text{Normal}\left(0,1\right)$, $\widehat{\Var}\left(r_{i} | D_i = d\right)\approx1$, and all $r_{i}$ are shrunk to zero. But if data are far from the null, then a large $r_{i}$ is a signal of large $\mu_{i}$---even if $r_{i}$ is found from searching tens of thousands of strategies, unguided by economic theory.

Figure \ref{fig:t-stat-1983} shows that equal-weighted
accounting strategies (upper left) are far from the null using data
from 1964 to 1983. Equal-weighted past return strategies (middle left)
also show a notable deviation. In contrast, the other strategy families
are quite close to the null. Indeed, for both families of ticker-based
strategies, the null is a very good fit for the data. 
\begin{center}
{[}Figure \ref{fig:t-stat-1983}, Distribution of t-stats
in 1983, about here{]} 
\par\end{center}

Accordingly, Equation (\ref{eq:shrink-ez}) implies that the strategies with strong actual performance will be found in equal-weighted accounting 
and equal-weighted past-return strategies. This intuition is consistent
with Panel A of Table \ref{tab:top_strat_desc}, which shows that
the vast majority of the best data-mined strategies come from these
families.

Compared to data available in 1983, all strategy families are closer
to the null using data from 1985-2004, as seen in Figure \ref{fig:t-stat-2004}.
All value-weighted families are very close to the null, implying that
predictability in large stocks is essentially gone. The long left
tail in equal-weighted past return strategies also disappears. Only
equal-weighted accounting strategies are visually far from the null.
These results imply that predictability is concentrated in the earlier
part of the sample. 
\begin{center}
{[}Figure \ref{fig:t-stat-2004}, Distribution of t-stats
in 2004, about here{]} 
\par\end{center}

The intuition in Figures \ref{fig:t-stat-1983} and \ref{fig:t-stat-2004}
is so simple that one might even skip the quasi-maximum likelihood
estimation. Just looking at these charts, and the distance between
the data and the null, one can already tell that predictability is
concentrated in small stocks, accounting data, and the earlier sample.
That is, one can already tell where predictability is concentrated,
if one understands the intuition in Equation (\ref{eq:shrink-ez}).

\section{Empirical Bayes Prediction Accuracy Across the Cross-Section\protect}\label{sec:accuracy}

This section takes a closer look at the EB predictions and accuracy.
We see when and where EB predictions are successful and when they
struggle.

\subsection{EB Prediction Accuracy 1983-2004\protect}\label{sec:oos-cross-pre2004}

To examine accuracy, we use out-of-sample portfolio sorts. For each
year and each strategy family, we form 20 portfolios by sorting strategies
into equal-sized groups based on the past 20 years of mean returns.
We then predict the mean returns for each portfolio by averaging the
EB predictions (Equation (\ref{eq:ephat-def})), which are also based
on the past 20 years of data. Finally, we form a portfolio that equally-weighs
strategies in each group and hold for one year (the ``out-of-sample''
periods).

Figure \ref{fig:xpred-1} shows the in-sample, predicted, and out-of-sample
returns for each portfolio, averaged over the out-of-sample periods from
1983 to 2004. For all six families, there are sizable in-sample returns
(dashed line) in the extreme in-sample groups. For accounting strategies,
in-sample returns are as extreme as -11\% per year. A naive read of
this result is that one can flip the long and short legs and find
+11\% returns out-of-sample. Past return strategies see a similar
$\pm10$ percent return in the extreme groups. Even ticker-based strategies
show in-sample long-short returns of up to 4 percent per year. 
\begin{center}
{[}Figure \ref{fig:xpred-1}, Empirical Bayes Predictions 1983-2004,
about here{]} 
\par\end{center}

However, the predicted returns are typically much closer to zero.
In fact, for both ticker-based strategy families, the predicted return
(solid line) is almost exactly zero for all 40 in-sample groups. This
result is intuitive given how close the ticker t-stats are to the
null of no predictability (Figure \ref{fig:t-stat-2004}).
This closeness implies that the extreme returns can be entirely accounted
for by luck, and so shrinkage should be 100\% (Equation (\ref{eq:shrink-ez})).
Significant shrinkage is also seen in value-weighted accounting strategies
(top right panel). Rigorous data mining recommends that the extreme
returns of around -8\% and +9\% (dashed line) be shrunk down to about
-3\% and +2 (solid line), respectively.

Rigorous mining predicts much higher returns in equal-weighted accounting
strategies (upper left panel). For these strategies, the predicted
returns are actually not far from the in-sample return. This result
is consistent with \citet{chen2020publication}, who find shrinkage
of only 12\% for published anomalies, which are largely equal-weighted
and based on accounting variables. Predictability is also seen in
both families of past return strategies.

These predictions are borne out in out-of-sample returns (markers
with error bars). The first group of EW accounting strategies returns
-8 percent per year out-of-sample from 1983-2004, almost exactly the
same as the EB prediction. Similar accuracy is seen throughout all
120 bins in Figure \ref{fig:xpred-1}.

These results show that rigorous data mining offers economic insights
that are difficult to derive from theory. While theories of slow information
diffusion may tell you that predictability is concentrated in small
stocks, accounting signals, and pre-2004 data, they are unlikely tell
you how much predictability there is. In contrast, empirical Bayes
provides quantitative, accurate estimates of the precise amount of
predictability.

\subsection{EB Prediction Accuracy 2004-2020\protect}\label{sec:oos-cross-post2004}

We split our OOS tests in the mid-2000s, motivated by the idea that
there was likely a structural break during this period due to the
rise of information technology (\citet{Chordia2014Have}). Comparing
the distribution of t-stats available in 1983 vs 2004 supports the
idea that the structure of financial markets changed (see Section
\ref{sec:best-intuition}). 
\begin{center}
{[}Figure \ref{fig:xpred-2}, Empirical Bayes Predictions 2004-2020,
about here{]} 
\par\end{center}

This structural change can be seen by comparing Figure \ref{fig:xpred-2}
(EB predictions 2004-2020) to Figure \ref{fig:xpred-1} (EB predictions
1983-2004). In all panels, the predicted returns shift closer to zero
post-2004. Most notably, the predictability that was present in past
return strategies pre-2004 is largely gone. Consistent with these
predictions, the past return portfolios show a flat or even negative
relationship between out-of-sample and in-sample returns post-2004.
A similar weakening of EB predictions and flattening of out-of-sample
returns is seen in the accounting VW family.

An exception to this pattern is the family of equal-weighted accounting
ratio strategies (top left). In this chart, the shrinkage is still
relatively small, with EB predictions implying returns as extreme
as -9 percent per year. This prediction and others in this panel miss
the mark: the out-of-sample returns are much closer to zero throughout
this panel.

This poor accuracy is natural given the fact that the estimations
use a rolling window consisting of the past 20 years of data. This
fixed window implies that, for much of the period 2004-2020, our estimates
rely on data from a time when accounting statements needed to be retrieved
by traditional (snail) mail for investors without special access to
the SEC reading room (\citet{bowles2023anomaly}).

This result implies an important role for economic theory: when structural
breaks occur, there is no way for data mining to provide a clear understanding
of the economy, no matter how rigorously the mining is done. Theory
is sometimes used this way in economics and finance, but this is typically
not the case. Instead, theory is typically used to understand patterns
found in long samples of data, spanning many decades. In our view,
the future of theory is bright for theorists who study structural
breaks, even in the era of big data. Indeed, a smart data miner armed
with theory might have understood the implications of the internet
for stock return predictability, and could perhaps have performed
much better than our theory-free EB mining process.

\section{Comparison with False Discovery Controls\protect}\label{sec:fdr}

Our main analysis corrects for data mining bias using empirical Bayes shrinkage, following \citet{chen2020publication}; \citet{chen2022zeroing}; and \citet{jensen2023there}. An alternative approach is to use false discovery controls, following \citet{harvey2016and}; \citet{barras2010false}; or \citet{chordia2020anomalies}. The ideal approach remains an unsettled question. Our dataset of 136,000 trading strategies provides a natural testing ground. 

We examine the following false discovery controls:
\begin{enumerate}
    \item \textbf{BY1.3 (1\%)}: \citet{harvey2016and} (HLZ) recommend using \citepos{benjamini2001control} Theorem 1.3 at the 1\% level. HLZ is likely the most influential paper on multiple testing in empirical asset pricing.

    \item \textbf{Storey (10\%)}: \citet{barras2010false}, which introduced false discovery methods to finance, study the \citet{storey2002direct} algorithm at the 10\% level. 

    \item \textbf{RW (5\%, 5\%)}: \citet{chordia2020anomalies} recommend combining \citepos{romano2007control} Algorithms 4.1 and 2.1. These algorithms require two parameters, both of which Chordia et al. set to 5\%.
\end{enumerate}
For each year and each strategy family, we apply these methods using the past 20 years of data to estimate a t-statistic hurdle. We then examine whether these hurdles are able to separate strategies with high out-of-sample returns from those with low out-of-sample returns. This structure is the same as in Section \ref{sec:accuracy}.

Figure \ref{fig:fd-control} shows the results. The vertical lines show the mean hurdle across all years. The markers show the mean out-of-sample returns of portfolios formed by equally weighting strategies, sorted into 20 groups based on the in-sample t-statistic. Groups of strategies that a false discovery control declares ``significant'' lie on the outside of the respective vertical lines. 

\begin{center}
    [Figure \ref{fig:fd-control}, False Discovery Controls, about here]
\end{center}

The BY1.3 (1\%) and RW (5\%, 5\%) methods miss out on the majority of portfolios with notable out-of-sample performance. Out of the 5 groups that have out-of-sample returns of at least 3\% per year, only 1 lies outside of the solid lines corresponding to BY1.3 (1\%). Only 3 of 5 lie outside the dot-dashed lines corresponding to RW (5\%, 5\%). The dashed line, corresponding to Storey (10\%), performs much better, capturing 4 of the 5 portfolios. Similar results are found using alternative parameter choices examined by HLZ; \citet{harvey2020false}; and \citet{barras2010false} (see Appendix Figure \ref{fig:fd-control-alt}).

Thus, Storey (10\%) provides an easy-to-compute alternative to empirical Bayes. However, Storey cannot provide bias-adjusted performance estimates that are naturally available from empirical Bayes. Overall, our results imply that Storey forms a strong first step for rigorous data mining, while empirical Bayes is recommended for more refined estimates. These results are broadly consistent with the statistics literature, which generally recommends Storey as a preliminary examination, while suggesting empirical Bayes for greater precision (e.g. \citealt{benjamini2010discovering}; \citealt{efron2012large}). 

We discuss this literature and the algorithms in more detail below.

\subsection{\citet{benjamini2001control} Theorem 1.3}\label{sec:fdr-hlz}

\citet{harvey2016and} (HLZ) recommend using \citepos{benjamini2001control}
Theorem 1.3. Several followups to the influential HLZ paper use this method, including \citet{harvey2020false,chordia2020anomalies}; and \citet{jensen2023there}.

We state the theorem number 1.3 because the bulk of the original paper focuses on Theorem 1.2. Indeed, \citet{benjamini2001control} describe Theorem 1.3 as ``very often unneeded, and yields too conservative of a procedure'' (page 1183). In his textbook on large scale inference, \citet{efron2012large} agrees, stating that the theorem represents a ``severe penalty'' and is ``not really necessary'' (section 4.2).  Moreover, the statistics literature uses the ``BY algorithm'' to refer to \citet{benjamini2005false}, which is an entirely different procedure (e.g. \citealt{efron2012large} Chapter 11.4).

BY1.3 begins by choosing a parameter $q^{\ast}$ and then solving
\begin{align}
h_{\text{HLZ},q^{\ast}} & \equiv\min_{h>0}\left\{ h:\left[\frac{\Pr\left(|Z|>h\right)}{\text{Share of \ensuremath{|t_{i}|>h}}}\right]\pi_{\text{BY1.3}}\le\text{\ensuremath{q^{\ast}}}\right\} \label{eq:by1.3hurdle}
\end{align}
where $t_i$ is the t-statistic for strategy $i$, $Z$ is a standard normal random variable, 
\begin{align}
\pi_{\text{BY1.3}} & \equiv\sum_{i=1}^{N}\frac{1}{i},
\label{eq:by1.3penalty}
\end{align}
and $N$ is the number of strategies in the year-family.  \citepos{benjamini2001control} Theorem 1.3 proves that this algorithm implies a false discovery rate
$\le q^{\ast}$. BY1.3 amounts to modifying the seminal \citet{benjamini1995controlling} algorithm with a constant factor, $\pi_{\text{BY1.3}}$. This modification makes the algorithm more conservative.

HLZ recommend this conservative approach, claiming  \citet{benjamini1995controlling} ``is only valid when the test statistics are independent or positively dependent'' (page 21). This statement is false. \citet{storey2001estimating} and \citet{storey2004strong} show validity under weak dependence assumptions (see also \citealt{chen2024most}). 

HLZ are also conservative in their choice of $q^{\ast}$. For their main results, they use  $q^{\ast}=1\%$ citing the fact that the ``significance level is subjective,'' though they also examine $q^{\ast}=5\%$ for robustness. In contrast, the statistics literature generally recommends $q^{\ast}=5\%$ or 10\% (e.g. \citealt{benjamini2010discovering}; \citealt{efron2012large}).  

Given this context, it is perhaps unsurprising that BY1.3 (1\%) fails to identify most out-of-sample performers in Figure \ref{fig:fd-control}. BY1.3 (5\%) performs somewhat better, identifying 2 out of 5 groups with out-of-sample returns of at least 3\% per year (Appendix Figure \ref{fig:fd-control-alt}). 

\subsection{\citepos{storey2002direct} FDR Control\protect}\label{sec:fdr-storey}

While HLZ recommend modifying \citet{benjamini1995controlling} to be more conservative, much of the statistics literature goes in the opposite direction, modifying \citet{benjamini1995controlling} to be more aggressive. In finance, \citet{barras2010false} take this approach.

Barras et al. recommend the \citet{storey2002direct}
algorithm, which can be written as
\begin{align}
h_{\text{Storey},q^{\ast}} & \equiv\min_{h>0}\left\{ h:\left[
    \frac{\Pr\left(|Z|>h\right)}{\text{Share of \ensuremath{|t_{i}|>h}}}
    \right]\pi_{\text{Storey}}\le\text{\ensuremath{q^{\ast}}}\right\} \label{eq:storey-1}
\end{align}
where $t_i$ is the t-statistic for strategy $i$, $Z$ is a standard normal random variable, 
\begin{align}
\pi_{\text{Storey}} & =\frac{\text{Share of \ensuremath{|t_{i}|\le1.0}}}{\Pr\left(|Z|\le 1.0\right)}=\frac{\text{Share of \ensuremath{|t_{i}|\le1.0}}}{0.68}\label{eq:storey-2}
\end{align}
and the cutoff of $1.0$ is selected for ease of interpretation. \citet{storey2002direct} proves that this algorithm implies a false discovery rate $\le q^{\ast}$ under independence assumptions, though \citet{storey2001estimating} and \citet{storey2004strong} extend this result to weak dependence.

Comparing Equations (\ref{eq:storey-1})-(\ref{eq:storey-2}) to the Equations (\ref{eq:by1.3hurdle})-(\ref{eq:by1.3penalty}), we see that the only difference is the constant factor, $\pi_{\text{BY1.3}}$ vs $\pi_{\text{Storey}}$. These constants are qualtitatively different: $\pi_{\text{BY1.3}} = \sum_{i=1}^{N}\frac{1}{i} \approx 0.6 + \log N \gg 1$ , while $\pi_{\text{Storey}}\le 1.0$. As shown in \citet{storey2002direct}, $\pi_{\text{Storey}}$ can be interpreted as an estimate of the probability that a strategy is null, which can be at most 1.0. Other statistics papers that recommend a constant that is at most 1.0 include \citet{benjamini2000adaptive, efron2001empirical,genovese2006false,benjamini2006adaptive}.

\citet{barras2010false} do not emphasize a particular choice of $q^{\ast}$, and instead examine values ranging from 5\% to 20\%. Figure \ref{fig:fd-control} uses $q^{\ast}=10\%$, because Barras et al. use 10\% in their illustrative examples. 

Once again, given the support from the statistics literature, it is perhaps unsurprising that Storey (10\%) and (20\%) perform well. Equations (\ref{eq:storey-1})-(\ref{eq:storey-2}) are easy to implement, making it a useful alternative to our empirical Bayes method.

However, there are two downsides to using Storey. A simple, symmetric testing algorithm like Equations (\ref{eq:storey-1})-(\ref{eq:storey-2}) does not handle skewed distributions well. This limitation may explain why Storey struggles to identify out-of-sample performers in past return strategies, which feature a long right tail (Figure \ref{fig:t-stat-1983}). The second is that Storey cannot provide bias-adjusted performance estimates. Such estimates are naturally available from a more general empirical Bayes method, and would provide clean connections with portfolio choice and asset pricing questions. 

\subsection{\citepos{romano2007control} FDP Risk Control}\label{sec:fdr-rw}

\citet{chordia2020anomalies} recommend combining \citepos{romano2007control} Algorithms 4.1 and 2.1, which we refer to as ``RW.'' This algorithm is a natural choice for asset pricing researchers, as its predecessor \citet{romano2005stepwise} is motivated by data mining for CAPM anomalies. Like HLZ's method, the Romano and Wolf methods have been used in influential asset pricing papers, including \citet{chordia2020anomalies}; \citet{engelberg2023do}; \citet{heath2023reusing}; and \citet{debodt2025competition}.

Unlike Storey and BY1.3, the statistics literature has relatively little discussion of the Romano and Wolf methods. Neither Romano and Wolf (\citeyear{romano2005stepwise}) nor Romano and Wolf (\citeyear{romano2007control}) is found in the textbooks \citet{efron2012large} and \citet{efron2016computer}. The two papers are also not found in the review articles on multiple testing \citet{benjamini2010discovering} and \citet{benjamini2020selective}. Thus, we provide some discussion here.

The goal of RW can be written as follows: find an $h$ that ensures
\begin{align}\label{eq:fdp-risk-control}
    \Pr(\FDP > p^\ast) \le q^\ast
\end{align}
where 
\begin{align}
    \FDP \equiv \frac{\text{
        Number of null strategies with $|t_i|>h$
        }}{\text{
            Number of strategies with $|t_i|>h$}}
\end{align}
and $p^\ast$ and $q^\ast$ are thresholds selected by the researcher. Null strategies are, typically, those with an actual performance of zero.

Figure \ref{fig:fdp-risk-demo} illustrates Equation (\ref{eq:fdp-risk-control}), by simulating one of our QML estimates many times. We run 2,000 simulations, each one consisting of 29,000 strategies. For simplicity, we assume all strategiesare independent.  The plot shows histograms of actual performance ($\mu_i$ in Equation \eqref{eq:r=mu+e}) for strategies that meet the hurdle $|t_i|>3.0$, where $h=3.0$ is selected for illustrative purposes. Using this chart, we can ask whether this $h=3.0$ hurdle achieves Equation \eqref{eq:fdp-risk-control}, and thus understand FDP risk control.

\begin{center}
    [Figure \ref{fig:fdp-risk-demo}, FDP Risk Control Illustration, about here]
\end{center}

Since the $h=3.0$ hurdle is quite stringent, the vast majority of strategies are non-null. However, there is still a risk that a strategy with $|t_i|>3.0$ has near-zero actual performance, as seen in the left tail of the histogram. The FDP characterizes this risk. It is, approximately, the share of strategies in the first bin.\footnote{More formally, one can consider the first bin to be an upper bound on the FDP (see \citealt{chen2024most}).}  On average, the share of strategies in this bin is about 5\% (bars), indicating that the FDR is approximately controlled at a 5\% level.

Even though the FDP is on average about 5\%, there is a risk that it is higher. This risk is seen in the lines of Figure \ref{fig:fdp-risk-demo}, which plot extreme order statistics across the 2,000 simulations. The 95th percentile line implies the FDP exceeds 7\% in 5\% of simulations. To achieve FDP risk control with $p^\ast=5\%$ and $q^\ast=5\%$, a $h>3.0$ is required. The RW method finds this $h$.

Thus, the RW method aims to control the tail risk of a tail risk. Such an algorithm is a natural choice if selecting a null strategy is catastrophic. In such a case, one may want to ensure not only that a null is highly improbable, but that the probability that a null is somewhat probable is also improbable. However, in the standard setting where the null is that the strategy has zero long-short return or zero alpha, then the RW method tends to imply extreme conservatism.

This conservatism leads to the results in Figure \ref{fig:fd-control} and Appendix Figure \ref{fig:fd-control-alt}. Choosing $p^\ast=0.05$ and $q^\ast=0.05$ or 0.10, as in \citet{chordia2020anomalies} and \citet{harvey2020evaluation}, leads to hurdles that many notable out-of-sample performers fail to clear.

The RW method is rather complex. It uses cluster bootstrap methods, involves testing all possible subsets of selected sets of strategies, iterating over many possible tests and sets. We describe our implementation in Appendix \ref{sec:app:rw-details} and provide code in our Github repo.

\section{Conclusion} 

We show that a solution to data mining bias is to mine data rigorously. We systematically search 136,000 long-short strategies and find out-of-sample performance comparable to academic research. Simply searching for strategies with the largest t-stats leads to publication-like out-of-sample performance, a fact we explain in a Bayesian model. While naive data mining leads to distorted performance estimates, empirical Bayes provides unbiased predictions in samples without structural breaks. The forecast errors around structural breaks suggest a role for theory in the era of big data.

This high-throughput method shows that returns are concentrated in accounting signals, small stocks, and pre-2004 periods, consistent with mispricing and slow information diffusion theories. While these results could potentially be gleaned from a deep read of the anomalies literature, our method provides a scientific method for documenting these stylized facts. We provide our data and code publicly, and hope others follow in using high-throughput methods.

Our out-of-sample tests offer an intuitive method for comparing multiple testing methods. We find that methods popular in finance would lead researchers to miss out on the majority of signals with notable out-of-sample performance. In contrast, methods recommended by the statistics literature perform well.

\newpage{}

\appendix

\renewcommand{\thefigure}{A.\arabic{figure}}
\renewcommand{\thetable}{A.\arabic{table}}

\section{Data Handling Details\protect}\label{sec:app:data}

\subsection{60,000 Accounting Ratio Strategies}

We examine 60,000 accounting ratio strategies constructed by \citet{chen2022peer}.
Inspired by \citet{yan2017fundamental}, Chen et al. construct 30,000
accounting ratio signals as follows. Let $X$ be one of 240 accounting
variables from Compustat (+ CRSP market equity) and $Y$ be one of
65 accounting of these 240 variables that is positive for at least
25\% of firms in 1963. Apply two transformations: $X/Y$ and $\Delta X/\text{lag}Y$
to get $240\times65\times2\approx30,000$ signals. Then form equal-weighted
and value-weighted long-short decile strategies, leading to 60,000
strategies.

These strategies are downloaded from Andrew Chen's website. We are
grateful to the others for making their data public.

\subsection{38,000 Past Return-Based Strategies}

Inspired by \citet{yan2017fundamental} , we construct past-return
strategies as follows: Choose 4 quarters out of the past 20 quarters.
Compute the first four central moments using the returns in these
quarters. This leads to $\binom{20}{4}\times4=19,380$ signals.

Add to this the return over any of the past 20 quarters, as well as
the mean return over the past 2 and past 3 quarters. This adds $20+2$
signals, for a total of $19,380+22=19,402$ signals.

Finally, form equal-weighted and value-weighted long-short decile
strategies.

We chose this approach, rather than the approach in \citet{yan2017fundamental}
for three reasons. The first is that we want to have a strategy list
that is comparable in length to the length of our accounting ratio
strategies. Yan and Zheng's method leads to ``only'' 4,080 signals.
The second is that, while Yan and Zheng's methods are inspired by
momentum and short-run reversal, we want to ensure that our methods
do not incorporate knowledge that would come from reading finance
publications. Last, we chose to reduce the amount of overlap across
the different signals, which should lead to better properties of our
EB estimator.

Earlier versions of our paper use Yan and Zheng's method and found
similar results. These results can be found at arxiv.org.

\subsection{38,000 Ticker-Based Strategies}

Inspired by \citet{harvey2017presidential}, we sort stocks into 20
groups based on the alphabetical order of the first ticker symbol.
We then long any two of those groups and short two. Repeat using the
2nd, 3rd, and 4th ticker symbols. This yields $\binom{20}{4}\times4=19,380$
long-short portfolios.

We chose not to follow \citet{harvey2017presidential}'s approach
in order to have a similar number of strategies as our accounting-based
strategies. Harvey's method leads to ``only'' 6,000 ticker-based
strategies.

Earlier versions of our paper used Harvey's method and found similar
results. These results can be found at arxiv.org.

\section{Theory and Estimation Details\protect}\label{sec:app:theory}

\subsection{Proof of Proposition \ref{prop:optimal-naive}}\label{sec:app:proof}

\begin{proof}
    Naive and EB mining will select the same strategies as long as, for $r_i > h$, $E(\mu_i|r_i, \SE_i, D_i)$ does not depend on  $\SE_i$ or $D_i$ and is strictly increasing in $r_i$.

    Condition \ref{cond:normal} implies that $\SE_i$ is constant. Thus $E(\mu_i|r_i, \SE_i, D_i)$ does not depend on  $\SE_i$.

    Condition \ref{cond:tail} implies that, for $r_i > h$, $E(\mu_i|r_i, \SE_i, D_i)$ does not depend on $D_i$. In this region,  Equation \ref{eq:tail-cond1} says that $D_i \in \mathcal{D}$. And then Equation \ref{eq:tail-cond2} says that if $D_i \in \mathcal{D}$, then $\mu_i|r_i, \SE_i$ does not depend on $D_i$.   Thus, $E(\mu_i|r_i, \SE_i, D_i)$ does not depend on  $\SE_i$ or $D_i$, and we can write, for $r_i > h$,
    \begin{align}
        E(\mu_\mathcal{D}|r_i, \SE_i, D_i ) = E(\mu_\mathcal{D}|r_i)
    \end{align}
    where $\mu_\mathcal{D}$ is a r.v. generated by $g_\SE\left(\cdot\right)$.

    Now we just need to show that for $r_i > h$, $ E(\mu_\mathcal{D}|r_i)$ is strictly increasing in $r_i$.  Tweedie's Formula (\citet{efron2011tweedie} Equation (2.8)) implies
\begin{align}
    E\left(\mu_\mathcal{D}|r_i\right)	
    &=r_i
    +\SE^2 \frac{d}{dr_i}\log f\left(r_i\right)\\
    \Var\left(\mu_\mathcal{D}|r_i\right)	
    &=
    \SE^2\left(
        1+\SE^2\frac{d^{2}}{dr_i^{2}}\log f\left(r_i\right)
    \right)
\end{align}
where $f(r_i)$ is the marginal density of $r_i$.   Differentiating the first equation with respect to $r_i$ and plugging in the second equation yields
\begin{align}
    \frac{d}{dr_i} E\left(\mu_\mathcal{D}|r_i\right)	
    &=  
    \frac{\Var\left(\mu_\mathcal{D}|r_i\right)}
    {\SE^2}   
    > 0
\end{align}   
where the inequality comes from the fact that $g_\SE\left(\cdot\right)$ has positive variance.

Thus, if $\Var(\mu_\mathcal{D}|r_i)$ is non-zero, then $ E\left(\mu_\mathcal{D}|r_i\right)$ is a strictly increasing function of $r_i$.
\end{proof}

\subsection{Estimation Details\protect}\label{sec:app:theory:estimation}

We construct the quasi-likelihood using the distr package in R (\citet{ruckdeschel2006s4})
and optimize using the BOBYQA algorithm in the nloptr package (\citet{NLopt}).
BOBYQA is a derivative-free bound-constrained optimization based on
quadratic approximations of the objective.

We also use distr to compute the prediction formula (Equation (\ref{eq:ephat-def})).
To ensure numerical stability, we split the integrals into many smaller
parts.

\subsection{RW Method Details}\label{sec:app:rw-details}

This approach is closely related to $k$-family-wise error rate ($k$-FWER) control, as the FDP can be thought of as the number of family-wise errors divided by the number of discoveries. 

We implement RW's Algorithm 4.1 as follows:
\begin{enumerate}
    \item Let $h$ be the t-stat threshold from applying RW Algorithm 2.1 to control the $k$-FWER at level $q^\ast$.
    \item Use bisection to find the largest $k$ such that 
    \begin{align}\label{eq:rw-p}
         \frac{k}
         {\left[\text{Number of } |t_{i}|>h\right]+1}
        > p^\ast
    \end{align} 
\end{enumerate}
This method modifies RW's Algorithm 4.1 to be more computationally efficient. Instead of testing every $k = 1,2,...$ until Condition (\ref{eq:rw-p}) is violated, we bisect to find this $k$ more quickly. As seen in the proof of Theorem 4.1 in RW, the Condition (\ref{eq:rw-p}) ensures $\Pr(\FDP > p^\ast) \le q^\ast$, while the sequence of $k$ examined does not matter.

We implement RW's Algorithm 2.1 as follows:
\begin{enumerate}[label=(\alph*)]
    \item Using all strategies, bootstrap 2,000 samples by demeaning returns at the strategy level and then re-sampling months with replacement. Let $|t^\ast_b|$ be the $k$th largest absolute t-stat across strategies in bootstrap $b$. Assign $h$ as the $(1-q^\ast)$ quantile of $|t^\ast_b|$ across the $B$ bootstraps.
    \item If $k=1$ then stop. If $\binom{\text{Number of strategies with $|t_i|>h$}}{k-1} > 100$ stop. Otherwise, repeat step (a) using the all possible sets of strategies that come from combining $k-1$ strategies with $|t_i| > h$ and all strategies with $|t_i| \le h$. Update $h$ to be the largest hurdle across all possible sets.
    \item Repeat step (b) until $h$ does not change.
\end{enumerate}
RW's Algorithm 2.1 does not have a stopping condition based on $\binom{\text{Number of strategies with $|t_i|>h$}}{k-1}$ but this condition is required to ensure that step (b) is computationally feasible. In fact, step (b) is typically infeasible in our setting, with tens of thousands of strategies. For example, if there are 1,000 strategies with $|t_i|>h$, and $k=3$, then there are $\binom{1000}{3}=166$ million possible sets to consider in step (b).  As seen in the proof of Theorem 2.1 in RW, imposing this stopping condition still ensures $k$-FWER control at level $q^\ast$.

\section{Robustness}\label{sec:app:results}

\clearpage\pagebreak 
\begin{figure}[h!] 
\caption{
    \textbf{False Discovery Controls: Robustness} We repeat the exercise in Figure \ref{fig:fd-control} using alternative parameter choices examined in \citet{harvey2016and} (for BY 1.3), \citet{barras2010false} (for Storey), and \citet{harvey2020false} (for RW).  \textbf{Interpretation:} As in Figure \ref{fig:fd-control}, the recommendations in Harvey et al. (2016) and Harvey et al. (2020) would lead one to miss most of the strategies with notable out-of-sample returns. Barras et al.'s recommendation performs much better.
}
\label{fig:fd-control-alt} \pdfbookmark{Figure A.1}{storey}\vspace{0.15in}

\centering \includegraphics[width=0.9\textwidth]{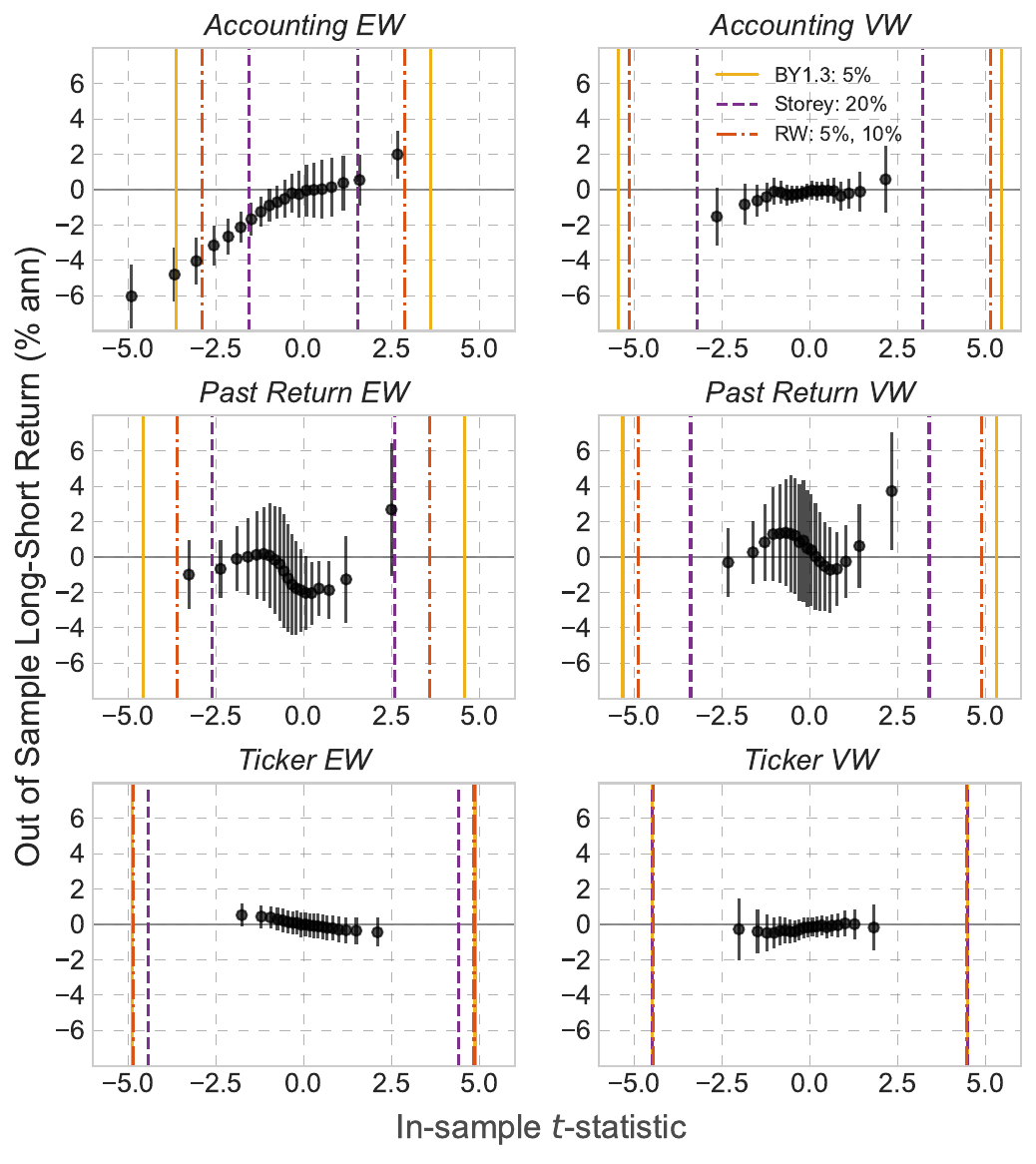}
\end{figure} 

\clearpage\pagebreak 
\begin{table} 
\caption{\textbf{Top 20 Strategies in 2003}}
\label{tab:top_strat_alt1} \pdfbookmark{Table A.1}{top_strat_desc}

\begin{singlespace}
\noindent We repeat Table \ref{tab:beststrats} Panel B using predicted Sharpe ratios from 1984-2003 and out-of-sample Sharpe ratios from 2004-2013. All strategies are equal-weighted. \textbf{Interpretation:} The list illustrates how the top strategies change over time. Shorting variables related to debt growth remains a common theme.
\end{singlespace}

\vspace{2ex}
 
\centering
\setlength{\tabcolsep}{0.5 ex}
\footnotesize
\begin{tabular}{cccll}
\toprule
Rank  & \makecell{Pred. SR \\ (ann)} & \makecell{OOS SR \\ (ann)} & \multicolumn{1}{c}{Signal Family} & Signal Name \\
\midrule
1     & 1.60  & 0.64  & Acct  & - LT Debt Issuance / Acc Depr, Depl \& Amort \\
2     & 1.53  & 1.27  & Acct  & - $\Delta$Interest \& Rel Exp / Lag(Tot Liabilities) \\
3     & 1.47  & 1.63  & Acct  & - $\Delta$Interest \& Rel Exp / Lag(Tot Assets) \\
4     & 1.46  & 0.90  & Acct  & - $\Delta$Interest \& Rel Exp / Lag(Parent SH Equity) \\
5     & 1.45  & 0.78  & Acct  & - $\Delta$Net Interest Paid / Lag(Parent SH Equity) \\
6     & 1.45  & 0.87  & Acct  & - $\Delta$Tot Liabilities / Lag(Tot Asset) \\
7     & 1.43  & 1.38  & Acct  & - $\Delta$Interest \& Rel Exp / Lag(Acc Depr, Depl \& Amort) \\
8     & 1.41  & 0.96  & Acct  & - $\Delta$Tot Liabilities / Lag(Curr Liabilities) \\
9     & 1.40  & 0.68  & Acct  & - $\Delta$Tot Liabilities / Lag(Other Curr Assets) \\
10    & 1.39  & 0.71  & Acct  & - $\Delta$Net Interest Paid / Lag(Comm Equity) \\
11    & 1.39  & 0.81  & Acct  & - $\Delta$Interest \& Rel Exp / Lag(Comm Equity) \\
12    & 1.38  & 1.57  & Acct  & - $\Delta$Interest \& Rel Exp / Lag(Invested Capital) \\
13    & 1.38  & 1.15  & Acct  & - Sale Comm \& Pref Stk / Cash \& ST Inv \\
14    & 1.38  & 0.64  & Acct  & - $\Delta$Tot Liabilities / Lag(Tot Liabilities) \\
15    & 1.38  & 0.17  & Acct  & - Mortg, Other Sec Debt / Acc Depr, Depl \& Amort \\
16    & 1.37  & 0.80  & Acct  & - $\Delta$Interest \& Rel Exp / Lag(Comm Equity Liq Val) \\
17    & 1.37  & 0.99  & Acct  & - $\Delta$Longterm Debt/ Lag(Acc Depr, Depl \& Amort) \\
18    & 1.37  & 0.75  & Acct  & - $\Delta$Tot Liabilities / Lag(Tot Curr Assets) \\
19    & 1.37  & 0.58  & Acct  & - $\Delta$Tot Longterm Debt / Lag(Other Curr Assets) \\
20    & 1.36  & 0.43  & Acct  & - $\Delta$Tot Longterm Debt / Lag(Com Equity Tangible) \\
\bottomrule
\end{tabular}%

\end{table} 

\clearpage\pagebreak 
\begin{table} 
\caption{\textbf{Top 20 Strategies in 2013}}
\label{tab:top_strat_alt2} \pdfbookmark{Table A.2}{top_strat_desc}

\begin{singlespace}
\noindent We repeat Table \ref{tab:beststrats} Panel B using predicted Sharpe ratios from 1994-2013 and out-of-sample Sharpe ratios from 2014-2020. All strategies are equal-weighted. \textbf{Interpretation:} Shorting variables related to debt growth remains a common theme.
\end{singlespace}

\vspace{2ex}
     
\centering
\setlength{\tabcolsep}{0.5 ex}
\footnotesize
\begin{tabular}{cccll}
\toprule
Rank  & \makecell{Pred. SR \\ (ann)} & \makecell{OOS SR \\ (ann)} & \multicolumn{1}{c}{Signal Family} & Signal Name \\
\midrule
1     & 1.40  & 0.49  & Acct  & - $\Delta$Interest \& Rel Exp / Lag(Total Assets) \\
2     & 1.35  & 0.29  & Acct  & - $\Delta$Interest \& Rel Exp / Lag(Invested Capital) \\
3     & 1.25  & -0.01 & Acct  & - $\Delta$Interest \& Rel Exp / Lag(Acc Depr, Depl \& Amort) \\
4     & 1.22  & 0.49  & Acct  & - Financing Actv, Net Cash / Cash \& ST Inv \\
5     & 1.20  & 0.52  & Acct  & - $\Delta$Interest \& Rel Exp / Lag(Market Val Equity) \\
6     & 1.18  & 0.34  & Acct  & - $\Delta$Interest \& Rel Exp / Lag(Tot Liabilities) \\
7     & 1.18  & 0.10  & Acct  & - $\Delta$Tot Liabilities / Lag(Total Assets) \\
8     & 1.17  & 0.32  & Acct  & - $\Delta$Net PPE / Lag(Dep \& Amort) \\
9     & 1.16  & 0.62  & Acct  & - $\Delta$Cost Goods Sold / Lag(Cost Goods Sold) \\
10    & 1.16  & -0.37 & Acct  & - $\Delta$Net Interest Paid / Lag(Acc Depr, Depl \& Amort) \\
11    & 1.15  & 0.60  & Acct  & - Sale Comm \& Pref Stk / Cash \& ST Inv \\
12    & 1.14  & 0.17  & Acct  & - $\Delta$Net Interest Paid / Lag(Total Assets) \\
13    & 1.14  & -0.37 & Acct  & - $\Delta$Interest \& Rel Exp / Lag(Gross PPE) \\
14    & 1.14  & -0.11 & Acct  & - $\Delta$Interest \& Rel Exp / Lag(Dep \& Amort) \\
15    & 1.13  & -0.32 & Acct  & - $\Delta$Tot Longterm Debt / Lag(Total Assets) \\
16    & 1.13  & 0.26  & Acct  & - $\Delta$Tot Liabilities / Lag(Curr Liabilities) \\
17    & 1.13  & -0.20 & Acct  & - $\Delta$Tot Longterm Debt / Lag(Acc Depr, Depl \& Amort) \\
18    & 1.13  & 0.19  & Acct  & - $\Delta$Interest \& Rel Exp / Lag(Parent SH Equity) \\
19    & 1.11  & 0.12  & Acct  & - $\Delta$Interest \& Rel Exp / Lag(Capital Exp) \\
20    & 1.11  & 0.42  & Acct  & - Financing Actv, Net Cash / Market Val Equity \\
\bottomrule
\end{tabular}%

\end{table}

\clearpage\pagebreak 

\printbibliography

@techreport{marrow2024real,
  title={Real-Time Discovery and Tracking of Return-Based Anomalies},
  author={Marrow, Benjamin and Nagel, Stefan},
  year={2024},
  institution={Working Paper}
}

@article{engelberg2023do,
  author = {Engelberg, Joseph and McLean, R. David and Pontiff, Jeffrey and Ringgenberg, Matthew C.},
  title = {Do Cross-Sectional Predictors Contain Systematic Information?},
  journal = {Journal of Financial and Quantitative Analysis},
  volume = {58},
  number = {3},
  pages = {1172--1201},
  year = {2023},
  doi = {10.2139/ssrn.3459229}
}

@article{heath2023reusing,
  author = {Heath, Davidson and Ringgenberg, Matthew C. and Samadi, Mehrdad and Werner, Ingrid M.},
  title = {Reusing Natural Experiments},
  journal = {Journal of Finance},
  volume = {78},
  number = {4},
  pages = {2329--2364},
  year = {2023},
  doi = {10.1111/jofi.13250}
}

@article{debodt2025competition,
  author = {de Bodt, Eric and Eckbo, B. Espen and Roll, Richard W.},
  title = {Competition Shocks, Rival Reactions, and Stock Return Comovement},
  journal = {Journal of Financial and Quantitative Analysis},
  note = {Published online 18 February 2025},
  year = {2025},
  doi = {10.1017/S0022109024000486}
}

@article{romano2005stepwise,
  title={Stepwise multiple testing as formalized data snooping},
  author={Romano, Joseph P and Wolf, Michael},
  journal={Econometrica},
  volume={73},
  number={4},
  pages={1237--1282},
  year={2005},
  publisher={Wiley Online Library}
}

@techreport{storey2001estimating,
  title={Estimating false discovery rates under dependence, with applications to DNA microarrays},
  author={Storey, John D and Tibshirani, Robert},
  year={2001},
  institution={Technical Report 2001-28, Department of Statistics, Stanford University}
}

@article{benjamini2010discovering,
  title={Discovering the false discovery rate},
  author={Benjamini, Yoav},
  journal={Journal of the Royal Statistical Society Series B: Statistical Methodology},
  volume={72},
  number={4},
  pages={405--416},
  year={2010},
  publisher={Oxford University Press}
}

@article{harvey2020evaluation,
  title={An evaluation of alternative multiple testing methods for finance applications},
  author={Harvey, Campbell R and Liu, Yan and Saretto, Alessio},
  journal={The Review of Asset Pricing Studies},
  volume={10},
  number={2},
  pages={199--248},
  year={2020},
  publisher={Oxford University Press}
}

@article{romano2007control,
  title={Control of Generalized Error Rates in Multiple Testing},
  author={Romano, Joseph P and Wolf, Michael},
  journal={The Annals of Statistics},
  pages={1378--1408},
  year={2007},
  publisher={JSTOR}
}

@article{lakonishok1994contrarian,
  title={Contrarian investment, extrapolation, and risk},
  author={Lakonishok, Josef and Shleifer, Andrei and Vishny, Robert W},
  journal={The journal of finance},
  volume={49},
  number={5},
  pages={1541--1578},
  year={1994},
  publisher={Wiley Online Library}
}

@article{peng2005learning,
  title={Learning with information capacity constraints},
  author={Peng, Lin},
  journal={Journal of Financial and Quantitative Analysis},
  volume={40},
  number={2},
  pages={307--329},
  year={2005},
  publisher={Cambridge University Press}
}

@techreport{didisheim2023complexity,
  title={Complexity in factor pricing models},
  author={Didisheim, Antoine and Ke, Shikun Barry and Kelly, Bryan T and Malamud, Semyon},
  year={2023},
  institution={National Bureau of Economic Research}
}

@article{kelly2024virtue,
  title={The virtue of complexity in return prediction},
  author={Kelly, Bryan and Malamud, Semyon and Zhou, Kangying},
  journal={The Journal of Finance},
  volume={79},
  number={1},
  pages={459--503},
  year={2024},
  publisher={Wiley Online Library}
}

@article{benjamini2006adaptive,
  title={Adaptive linear step-up procedures that control the false discovery rate},
  author={Benjamini, Yoav and Krieger, Abba M and Yekutieli, Daniel},
  journal={Biometrika},
  volume={93},
  number={3},
  pages={491--507},
  year={2006},
  publisher={Oxford University Press}
}

@article{benjamini2000adaptive,
  title={On the adaptive control of the false discovery rate in multiple testing with independent statistics},
  author={Benjamini, Yoav and Hochberg, Yosef},
  journal={Journal of educational and Behavioral Statistics},
  volume={25},
  number={1},
  pages={60--83},
  year={2000},
  publisher={SAGE Publications Sage CA: Los Angeles, CA}
}

@article{efron2001empirical,
  title={Empirical Bayes analysis of a microarray experiment},
  author={Efron, Bradley and Tibshirani, Robert and Storey, John D and Tusher, Virginia},
  journal={Journal of the American statistical association},
  volume={96},
  number={456},
  pages={1151--1160},
  year={2001},
  publisher={Taylor \& Francis}
}

@misc{NLopt,
  title = {The {NLopt} nonlinear-optimization package},
  author = {Steven G. Johnson},
  year = {2007},
  howpublished = {\url{https://github.com/stevengj/nlopt}}
}

@Article{ruckdeschel2006s4,
  title = {S4 Classes for Distributions},
  author = {P. Ruckdeschel and M. Kohl and T. Stabla and F.
    Camphausen},
  language = {English},
  journal = {R News},
  year = {2006},
  volume = {6},
  number = {2},
  pages = {2--6},
  month = {May},
}

@book{efron2016computer,
  title={Computer age statistical inference: Data mining, inference and prediction},
  author={Efron, B and Hastie, T},
  year={2016},
  publisher={Cambridge: Cambridge University Press}
}

@article{kim2021causal,
  title={Causal Effect of Information Costs on Asset Pricing Anomalies},
  author={Kim, Yong Hyuck and Ivkovich, Zoran and Muravyev, Dmitriy},
  journal={Available at SSRN 3921785},
  year={2021}
}

@article{yang2021high,
  title={High-throughput methods in the discovery and study of biomaterials and materiobiology},
  author={Yang, Liangliang and Pijuan-Galito, Sara and Rho, Hoon Suk and Vasilevich, Aliaksei S and Eren, Aysegul Dede and Ge, Lu and Habibovic, Pamela and Alexander, Morgan R and de Boer, Jan and Carlier, Aurelie and others},
  journal={Chemical reviews},
  volume={121},
  number={8},
  pages={4561--4677},
  year={2021},
  publisher={ACS Publications}
}

@article{storey2002direct,
  title={A direct approach to false discovery rates},
  author={Storey, John D},
  journal={Journal of the Royal Statistical Society Series B: Statistical Methodology},
  volume={64},
  number={3},
  pages={479--498},
  year={2002},
  publisher={Oxford University Press}
}

@article{chen2022peer,
  title={Peer-reviewed theory does not help predict the cross-section of stock returns},
  author={Chen, Andrew Y and Lopez-Lira, Alejandro and Zimmermann, Tom},
  journal={arXiv preprint arXiv:2212.10317},
  year={2022}
}

@article{bowles2023anomaly,
  title={Anomaly time},
  author={Bowles, Boone and Reed, Adam V and Ringgenberg, Matthew C and Thornock, Jacob R},
  journal={Available at SSRN 3069026},
  year={2023}
}

@article{efron1973stein,
  title={Stein's estimation rule and its competitors-an empirical Bayes approach},
  author={Efron, Bradley and Morris, Carl},
  journal={Journal of the American Statistical Association},
  volume={68},
  number={341},
  pages={117--130},
  year={1973},
  publisher={Taylor \& Francis}
}

@article{robbins1956empirical,
  title={An Empirical Bayes Approach to Statistics},
  author={Robbins, Herbert},
  journal={Proceedings of the Third Berkeley Symposium on Mathematical Statistics and Probability, Volume 1: Contributions to the Theory of Statistics},
  volume={3.1},
  year={1956},
  publisher={JSTOR}
}

@article{efron2011tweedie,
  title={Tweedie's formula and selection bias},
  author={Efron, Bradley},
  journal={Journal of the American Statistical Association},
  volume={106},
  number={496},
  pages={1602--1614},
  year={2011},
  publisher={Taylor \& Francis}
}

@article{chen2022zeroing,
  title={Zeroing in on the Expected Returns of Anomalies},
  author={Chen, Andrew Y and Velikov, Mihail},
  journal={Journal of Financial and Quantitative Analysis},
  year={2022}
}

@article{benjamini2005false,
  title={False discovery rate--adjusted multiple confidence intervals for selected parameters},
  author={Benjamini, Yoav and Yekutieli, Daniel},
  journal={Journal of the American Statistical Association},
  volume={100},
  number={469},
  pages={71--81},
  year={2005},
  publisher={Taylor \& Francis}
}

@article{genovese2006false,
  title={False discovery control with p-value weighting},
  author={Genovese, Christopher R and Roeder, Kathryn and Wasserman, Larry},
  journal={Biometrika},
  volume={93},
  number={3},
  pages={509--524},
  year={2006},
  publisher={Oxford University Press}
}

@article{benjamini2020selective,
  title={Selective inference: The silent killer of replicability},
  author={Benjamini, Yoav},
  year={2020},
  publisher={PubPub}
}

@article{cochrane2005risk,
  title={The risk and return of venture capital},
  author={Cochrane, John H},
  journal={Journal of financial economics},
  volume={75},
  number={1},
  pages={3--52},
  year={2005},
  publisher={Elsevier}
}

@article{fisher1925statistical,
  title={Statistical methods for research workers.},
  author={Fisher, RA},
  year={1925},
  publisher={Stechert}
}

@article{fama1993common,
  title={Common risk factors in the returns on stocks and bonds},
  author={Fama, Eugene F and French, Kenneth R},
  journal={Journal of financial economics},
  volume={33},
  number={1},
  pages={3--56},
  year={1993},
  publisher={Elsevier}
}

@article{fama1992cross,
  title={The cross-section of expected stock returns},
  author={Fama, Eugene F and French, Kenneth R},
  journal={the Journal of Finance},
  volume={47},
  number={2},
  pages={427--465},
  year={1992},
  publisher={Wiley Online Library}
}

@article{harvey2020false,
  title={False (and missed) discoveries in financial economics},
  author={Harvey, Campbell R and Liu, Yan},
  journal={The Journal of Finance},
  volume={75},
  number={5},
  pages={2503--2553},
  year={2020},
  publisher={Wiley Online Library}
}

@article{barras2010false,
  title={False discoveries in mutual fund performance: Measuring luck in estimated alphas},
  author={Barras, Laurent and Scaillet, Olivier and Wermers, Russ},
  journal={The journal of finance},
  volume={65},
  number={1},
  pages={179--216},
  year={2010},
  publisher={Wiley Online Library}
}

@article{chen2024t,
  title={Do t-Statistic Hurdles Need to be Raised?},
  author={Chen, Andrew Y},
  journal={Management Science},
  year={2024},
  publisher={INFORMS}
}

@article{storey2004strong,
  title={Strong control, conservative point estimation and simultaneous conservative consistency of false discovery rates: a unified approach},
  author={Storey, John D and Taylor, Jonathan E and Siegmund, David},
  journal={Journal of the Royal Statistical Society: Series B (Statistical Methodology)},
  volume={66},
  number={1},
  pages={187--205},
  year={2004},
  publisher={Wiley Online Library}
}

@article{jegadeesh1993returns,
  title={Returns to buying winners and selling losers: Implications for stock market efficiency},
  author={Jegadeesh, Narasimhan and Titman, Sheridan},
  journal={The Journal of finance},
  volume={48},
  number={1},
  pages={65--91},
  year={1993},
  publisher={Wiley Online Library}
}

@article{benjamini2001control,
  title={The control of the false discovery rate in multiple testing under dependency},
  author={Benjamini, Yoav and Yekutieli, Daniel},
  journal={Annals of statistics},
  pages={1165--1188},
  year={2001},
  publisher={JSTOR}
}

@article{chen2024most,
  title={Most claimed statistical findings in cross-sectional return predictability are likely true},
  author={Chen, Andrew Y},
  journal={arXiv preprint arXiv:2206.15365},
  year={2024}
}

@article{harvey2016and,
  title={... and the cross-section of expected returns},
  author={Harvey, Campbell R and Liu, Yan and Zhu, Heqing},
  journal={The Review of Financial Studies},
  volume={29},
  number={1},
  pages={5--68},
  year={2016},
  publisher={Oxford University Press}
}

@article{benjamini1995controlling,
  title={Controlling the false discovery rate: a practical and powerful approach to multiple testing},
  author={Benjamini, Yoav and Hochberg, Yosef},
  journal={Journal of the Royal statistical society: series B (Methodological)},
  volume={57},
  number={1},
  pages={289--300},
  year={1995},
  publisher={Wiley Online Library}
}

@article{chen2020publication,
  title={Publication bias and the cross-section of stock returns},
  author={Chen, Andrew Y and Zimmermann, Tom},
  journal={The Review of Asset Pricing Studies},
  volume={10},
  number={2},
  pages={249--289},
  year={2020},
  publisher={Oxford University Press}
}

@article{yan2017fundamental,
  title={Fundamental analysis and the cross-section of stock returns: A data-mining approach},
  author={Yan, Xuemin Sterling and Zheng, Lingling},
  journal={The Review of Financial Studies},
  volume={30},
  number={4},
  pages={1382--1423},
  year={2017},
  publisher={Oxford University Press}
}

@article{ChenZimmermann2021,
  title={Open Source Cross Sectional Asset Pricing},
  author={Chen, Andrew Y. and Tom Zimmermann},
  journal={Critical Finance Review},
  year={2022}
}

@book{efron2012large,
  title={Large-scale inference: empirical Bayes methods for estimation, testing, and prediction},
  author={Efron, Bradley},
  volume={1},
  year={2012},
  publisher={Cambridge University Press}
}

@article{harvey2017presidential,
  title={Presidential address: The scientific outlook in financial economics},
  author={Harvey, Campbell R},
  journal={The Journal of Finance},
  volume={72},
  number={4},
  pages={1399--1440},
  year={2017},
  publisher={Wiley Online Library}
}

@article{Chordia2014Have,
  title={Have capital market anomalies attenuated in the recent era of high liquidity and trading activity?},
  author={Chordia, Tarun and Subrahmanyam, Avanidhar and Tong, Qing},
  journal={Journal of Accounting and Economics},
  volume={58},
  number={1},
  pages={41--58},
  year={2014},
  publisher={Elsevier}
}

@article{chordia2020anomalies,
  title={Anomalies and false rejections},
  author={Chordia, Tarun and Goyal, Amit and Saretto, Alessio},
  journal={The Review of Financial Studies},
  volume={33},
  number={5},
  pages={2134--2179},
  year={2020},
  publisher={Oxford University Press}
}

@article{jensen2023there,
  title={Is there a replication crisis in finance?},
  author={Jensen, Theis Ingerslev and Kelly, Bryan and Pedersen, Lasse Heje},
  journal={The Journal of Finance},
  volume={78},
  number={5},
  pages={2465--2518},
  year={2023},
  publisher={Wiley Online Library}
}

@article{Mclean2016Does,
  title={Does academic research destroy stock return predictability?},
  author={McLean, R David and Pontiff, Jeffrey},
  journal={The Journal of Finance},
  volume={71},
  number={1},
  pages={5--32},
  year={2016},
  publisher={Wiley Online Library}
}

\pagebreak

\setcounter{table}{0} 
\global\long\def\thetable{\arabic{table}}%
 \setcounter{figure}{0} 
\global\long\def\thefigure{\arabic{figure}}%

\section*{Figures}

\begin{figure}[h!]
\caption{\textbf{Cumulative Long-Short Returns from Rigorous Data-Mining. }Each year, we sign strategies to have positive returns and form portfolios that equal-weight the top $X\%$ of predicted Sharpe ratios.  ``EB Mining'' uses Equation (\ref{eq:ephat-def}) while ``Naive Mining'' uses the standard calculation. We hold for one year and repeat.\textbf{ Interpretation:
} Like published strategies, data-mined strategies show little cyclicality. Both published and data-mined strategies experience a break around the early-2000s, around the time when internet access became widespread.}
\label{fig:beststrats-cret} \pdfbookmark{Figure 1}{cret}\vspace{0.15in}

\centering \includegraphics[width=0.9\textwidth]{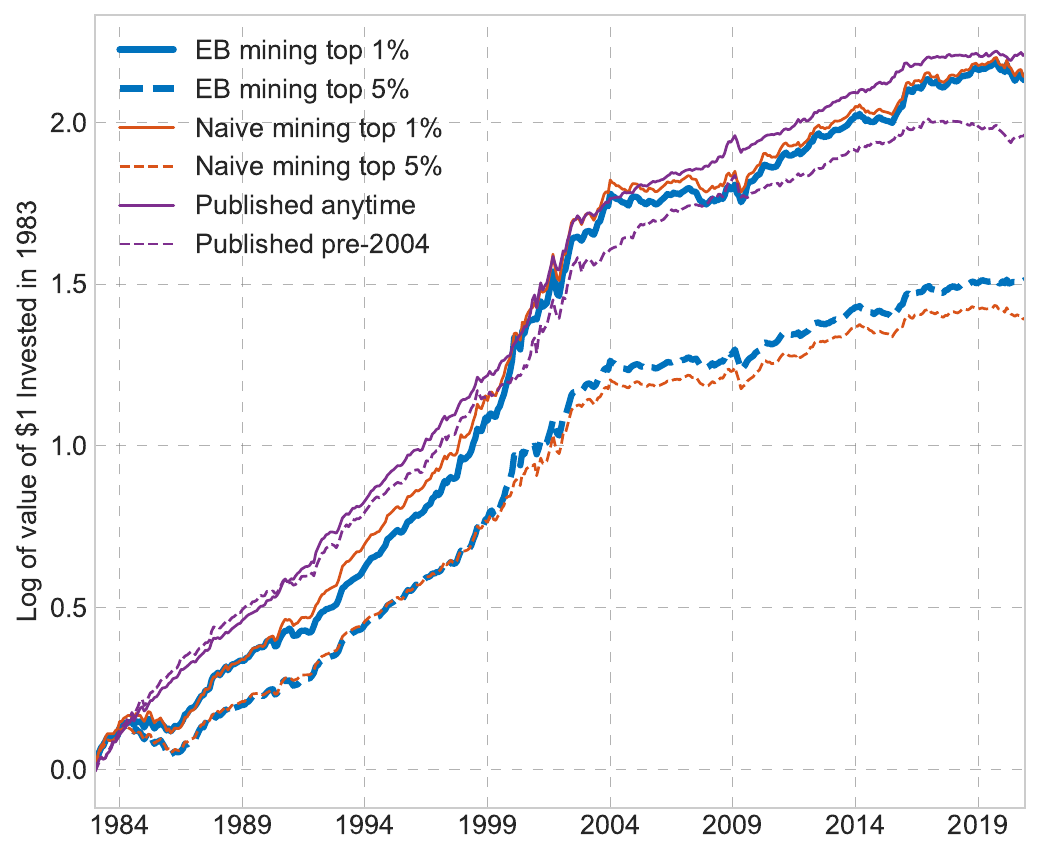} 
\end{figure}

\clearpage\pagebreak 
\begin{figure}[h!]
\caption{\textbf{Distribution of t-stats from long-short deciles strategies:
1983. }``Data'' are t-stats testing the null of expected return
= 0 from 1964-1983 for 136,000 trading strategies (Table \ref{tab:data-descrip}).
``Model'' is Equations (\ref{eq:t=theta+delta})-(\ref{eq:theta~mixnorm}).
``Null'' is a standard normal. ``EW'' and ``VW'' are equal-
and value-weighting, respectively. \textbf{Interpretation: }Equal-weighted
accounting and equal-weighted past return strategies are far from
the null, indicating true predictability. The models fit the data
well.}
\label{fig:t-stat-1983} \pdfbookmark{Figure 2}{t-stat dist}\vspace{0.15in}

\centering \includegraphics[width=0.9\textwidth]{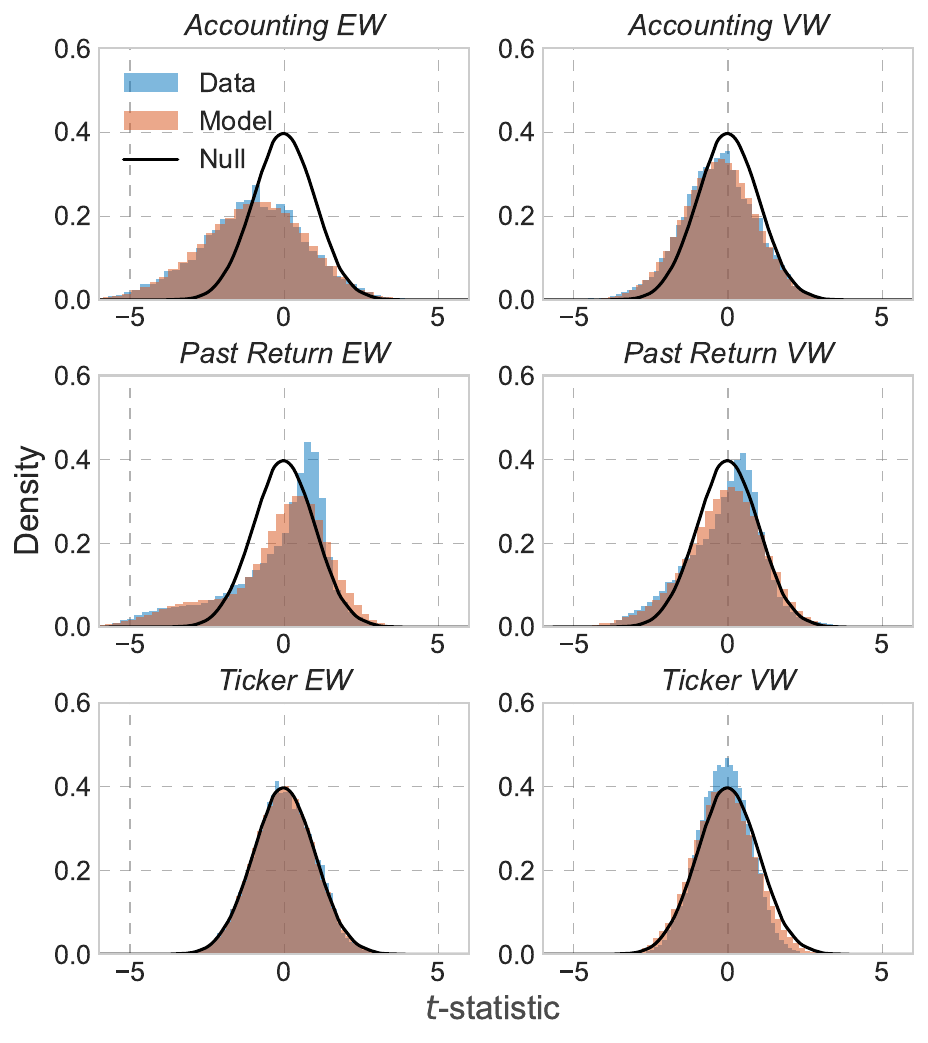} 
\end{figure}

\newpage\clearpage{}

\begin{figure}[h!]
\caption{\textbf{Distribution of t-stats from long-short deciles strategies:
2004. }``Data'' are t-stats testing the null of expected return
= 0 from 1985-2004 for 136,000 trading strategies (Table \ref{tab:data-descrip}).
``Model'' is Equations (\ref{eq:t=theta+delta})-(\ref{eq:theta~mixnorm}).
``Null'' is a standard normal. ``EW'' and ``VW'' are equal-
and value-weighting, respectively. \textbf{Interpretation: }Compared
to 1983 (Figure \ref{fig:t-stat-1983}), t-stats from
2004 are much closer to the null, indicating diminished predictability.
Equal-weighted accounting strategies still show true predictability,
however.}
\label{fig:t-stat-2004} \pdfbookmark{Figure 3}{t-stat dist 2}\vspace{0.15in}

\centering \includegraphics[width=0.9\textwidth]{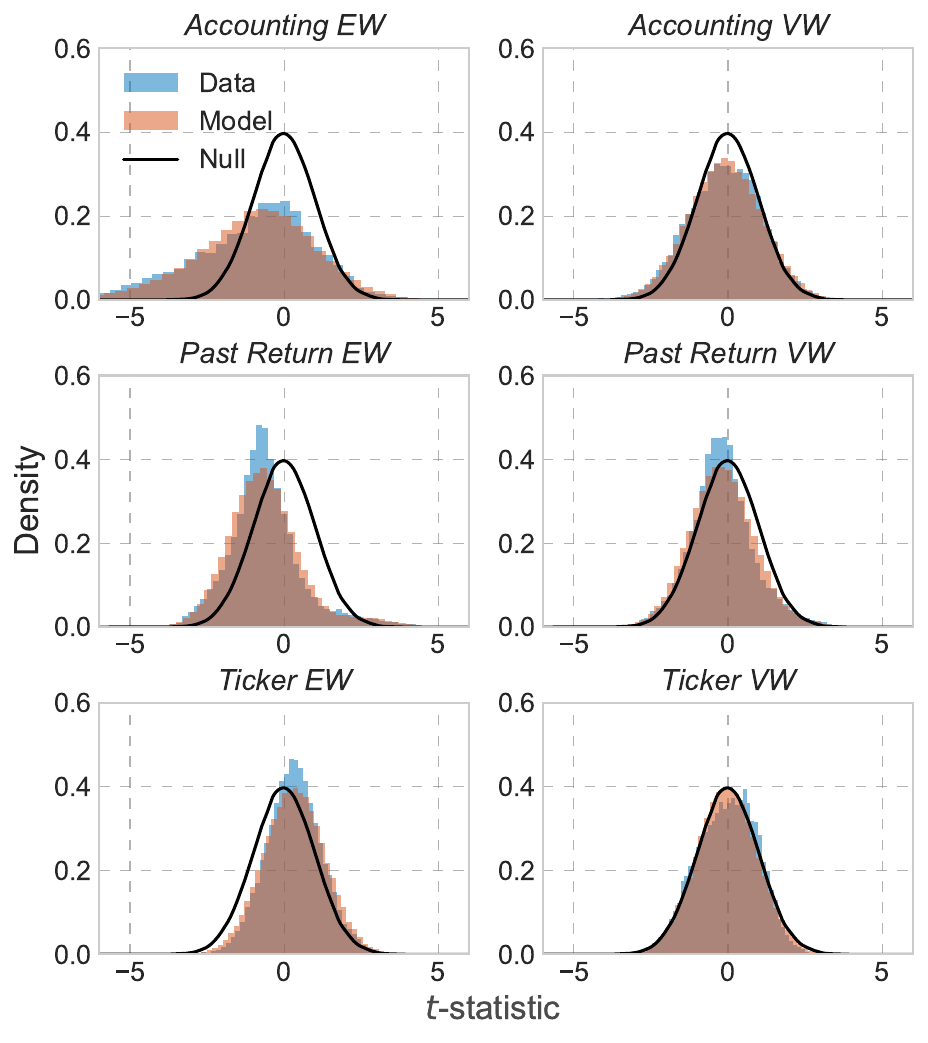} 
\end{figure}

\newpage\clearpage{}

\begin{figure}[h!]
\caption{\textbf{Empirical Bayes Predictions and Out-of-Sample Returns: 1983-2004.
}For each year and each family of strategies, we sort strategies into
20 groups based on the past 20 years of returns (``In-Samp'') and
predict returns using Bayes rule (Equation (\ref{eq:luck-sketch}),
``Predicted''). We form equal-weighted portfolios of strategies
in each group and hold for one year (``OOS,'' error bars are two
standard errors). \textbf{Interpretation: }Pre-2004, empirical Bayes
shrinkage provides accurate forecasts of out-of-sample returns, unlike
using the naive rule of in-sample return = out-of-sample return. Rigorous
data mining removes data mining bias.}
\label{fig:xpred-1} \pdfbookmark{Figure 4}{shrinkage vs oos 1}\vspace{0.15in}

\centering \includegraphics[width=1\textwidth]{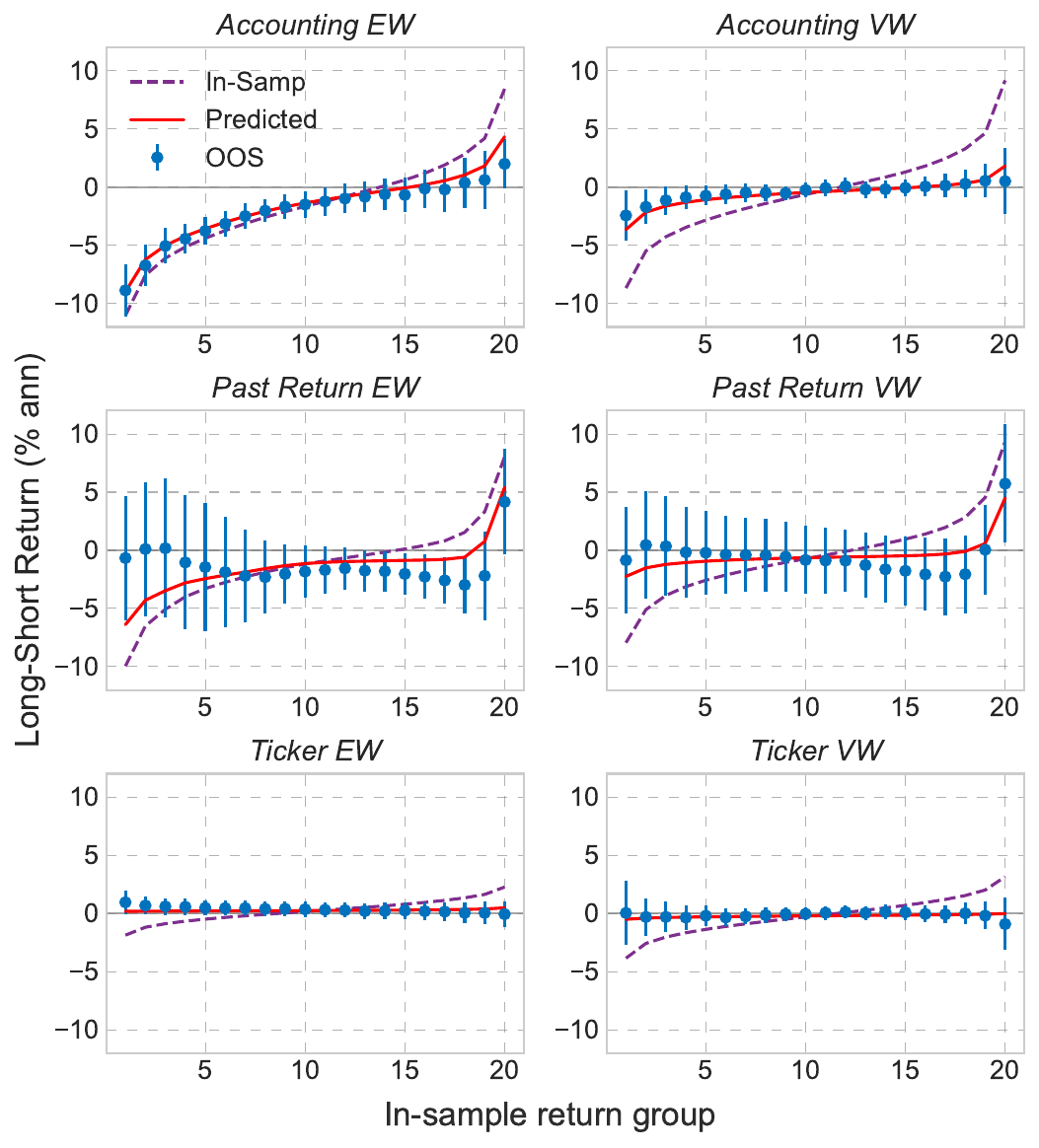} 
\end{figure}

\newpage\clearpage{} 
\begin{figure}[h!]
\caption{\textbf{Empirical Bayes Predictions and Out-of-Sample Returns: 2004-2020.
}For each year and each family of strategies, we sort strategies into
20 groups based on the past 20 years of returns (``In-Samp'') and
predict returns using Bayes rule (Equation (\ref{eq:luck-sketch}),
``Predicted''). We form equal-weighted portfolios of strategies
in each group and hold for one year (``OOS,'' error bars are two
standard errors). \textbf{Interpretation: }Compared with pre-2004
(Figure \ref{fig:xpred-1}), post-2004 predicted returns are closer
to zero. Out-of-sample returns are even closer to zero, consistent
with a structural break in predictability around 2004.}
\label{fig:xpred-2} \pdfbookmark{Figure 5}{shrinkage vs oos 2}\vspace{0.15in}

\centering \includegraphics[width=1\textwidth]{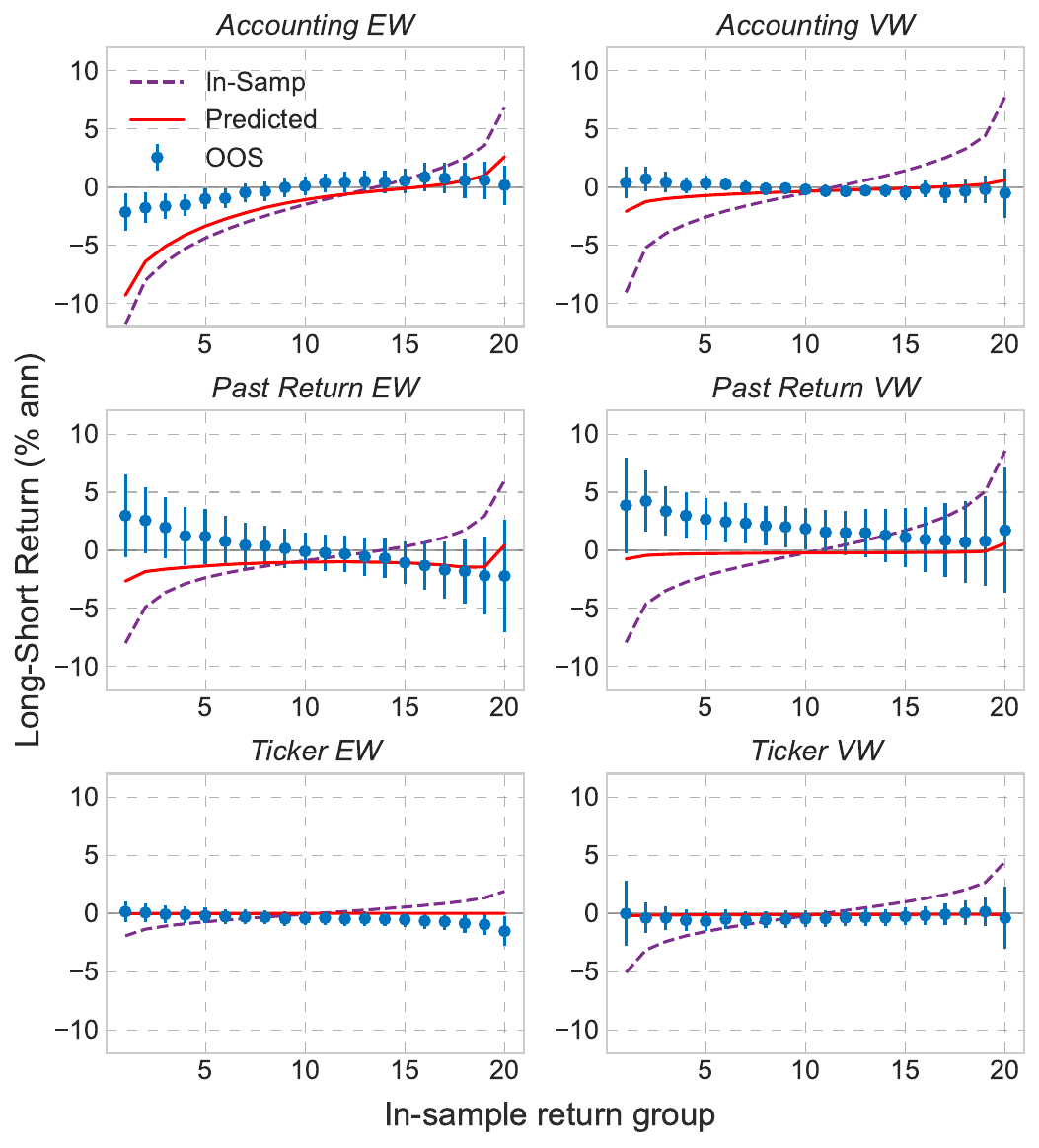} 
\end{figure}

\newpage\clearpage{} 
\begin{figure}[h!]
\caption{\textbf{False Discovery Controls} For each year and each strategy family, we calculate t-stat hurdles (vertical lines) using the recommendations of \citet{harvey2016and} (BY 1.3: 1\%); \citet{barras2010false} (Storey: 10\%); and \citet{chordia2020anomalies} (RW: 5\%, 5\%). We compare with out-of-sample returns of strategies sorted into 20 bins based on in-sample t-statistics (markers). Hurdles, in-sample t-stats, and out-of-sample returns are calculated each year from 1983-2020, and then averaged across years. Error bars are two standard errors.\textbf{ Interpretation:} Following the recommendations of Harvey et al. and Chordia et al. would lead one to miss most of the strategies with notable out-of-sample returns. The recommendation of Barras et al. performs much better.}
\label{fig:fd-control} 
\pdfbookmark{Figure 6}{by1.3}\vspace{0.15in}

\centering \includegraphics[width=1.00\textwidth]{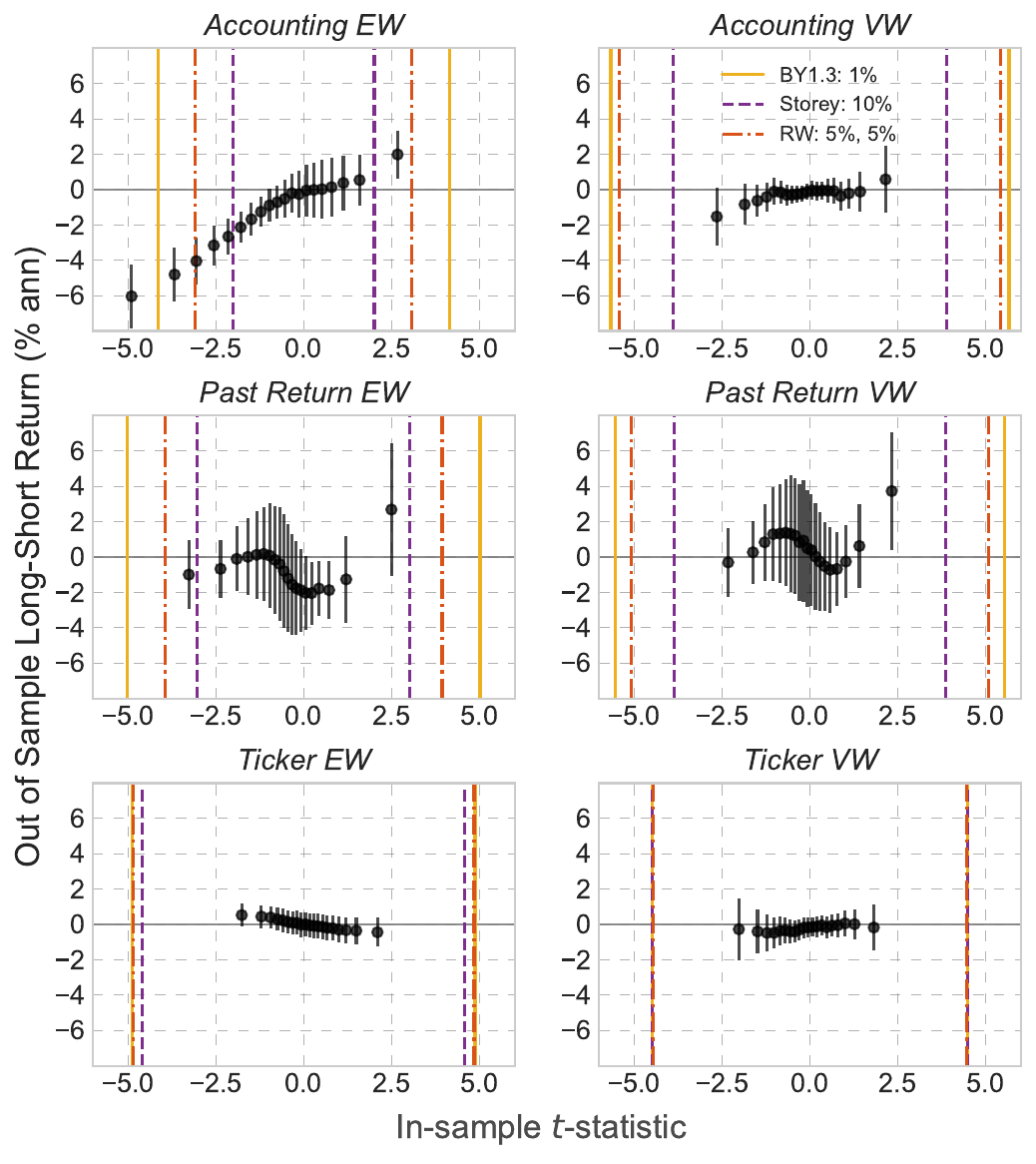}
\end{figure}

\newpage\clearpage{} 
\begin{figure}[h!]
\caption{\textbf{FDP Risk Control Illustration.} Using the QML estimates for value-weighted accounting strategies based on data from 1964-1983, we run 2,000 simulations of 29,000 strategies, filter for $|t_i| > 3.0$, and then calculate histogram counts. For simplicity, we assume all signals are independent. The plot shows various statistics for each histogram bin, calculated across simulations. The FDP is approximately the share of strategies in the first bin. \textbf{Interpretation:} On average, the FDP is 5\%, meaning the FDR is approximately controlled with a hurdle of 3.0. However, there is a risk that $\FDP> 5\%$, and thus a hurdle $>$ 3.0 is needed  to ensure $\Pr(\FDP > 5\%) < 5\%$. Since the FDP itself measures a left tail, the RW method aims to control the tail risk of a tail risk. This consevatism only makes sense if selecting a null strategy is catastrophic.}
\label{fig:fdp-risk-demo} \pdfbookmark{Figure 7}{fdp-risk-demo}\vspace{0.15in}

\centering \includegraphics[width=0.9\textwidth]{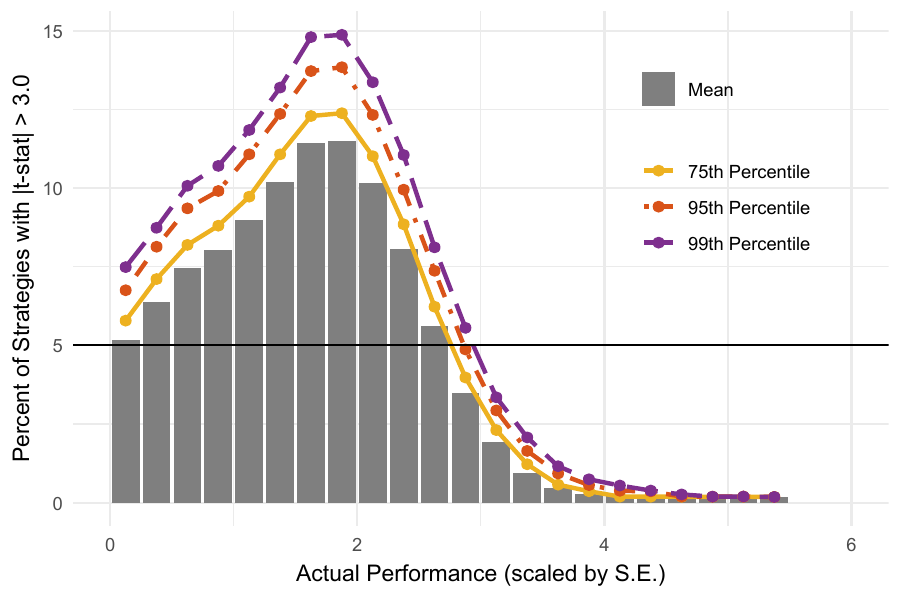}
\end{figure}

\newpage

\newgeometry{top=0.5in, bottom=0.5in} 
\section*{Tables}

\begin{table}[h]
\caption{\textbf{Overview of 136,000 Long-Short Strategies}}
\label{tab:data-descrip} \pdfbookmark{Table 1}{data-descrip}

\begin{singlespace} 
\noindent Table describes the 136,192 strategies used throughout the
paper. Data and code for these strategies are posted publicly. \textbf{Interpretation:
}Unlike datasets of published strategies (e.g. \citet{ChenZimmermann2021}),
these strategies are arguably constructed without data-mining bias. 
\end{singlespace}

\begin{centering}
\vspace{-0ex}
 
\par\end{centering}
\centering{}\setlength{\tabcolsep}{3ex} 
\centering{}{\small
\begin{tabular}{ccccc}
\toprule
\multicolumn{5}{c}{Panel A: Accounting Strategies} \\
\midrule
\multicolumn{5}{p{35.275em}}{Description: Make ratios from 242 accounting variables by (1) dividing one variable by another and (2) taking first differences and then dividing.  Long / short the extreme deciles.  Data is from \citet{chen2022peer}.} \\
\midrule
      & \multirow{2}[4]{*}{\# strategies} & \multicolumn{3}{c}{Mean Return (\% ann)} \\
\cmidrule{3-5}      &       & 5 pctile & 50 pctile & 95 pctile \\
\cmidrule{2-5}EW    & 29,314 & -7.0    & -1.1  & 3.7 \\
VW    & 29,314 & -4.5  & -0.4  & 3.9 \\
\midrule
\multicolumn{5}{c}{Panel B: Past Return Strategies} \\
\midrule
\multicolumn{5}{p{35.275em}}{Description: Choose 4 quarters out of the past 20 and compute one of the first four central moments, yielding $\binom{20}{4}\times 4=19,380$ signals. Add the return over any of the past 20 quarters and mean returns over the past 2 and past 3 quarters to arrive at 19,402 signals. Long / short the extreme deciles.} \\
\midrule
      & \multirow{2}[4]{*}{\# strategies} & \multicolumn{3}{c}{Mean Return (\% ann)} \\
\cmidrule{3-5}      &       & 5 pctile & 50 pctile & 95 pctile \\
\cmidrule{2-5}EW    & 19,402 & -5.3  & -0.4  & 2.1 \\
VW    & 19,402 & -3.4  & 0.1   & 4.3 \\
\midrule
\multicolumn{5}{c}{Panel C: Ticker Strategies} \\
\midrule
\multicolumn{5}{p{35.275em}}{Description: Sort stocks into 20 groups based on alphabetical order of the first ticker symbol. Long two of those groups and short two. Repeat using the 2nd, 3rd, and 4th ticker symbols. This yields $\binom{20}{4}\times 4 = 19,380$ long-short portfolios.} \\
\midrule
      & \multirow{2}[4]{*}{\# strategies} & \multicolumn{3}{c}{Mean Return (\% ann)} \\
\cmidrule{3-5}      &       & 5 pctile & 50 pctile & 95 pctile \\
\cmidrule{2-5}EW    & 19,380 & -0.9  & 0.0   & 0.8 \\
VW    & 19,380 & -2.2  & -0.2  & 1.6 \\
\bottomrule
\end{tabular}%
 }{\small\par}
\end{table}
\restoregeometry 

\newpage
\begin{table}
\caption{\textbf{Returns of Data-Mined Long-Short Portfolios}}
\label{tab:beststrats} \pdfbookmark{Table 2}{bestret}

\begin{singlespace}
\noindent Each year, we sign strategies to have positive predicted returns and form portfolios that equally-weights the top $X\%$ of strategies based on their Sharpe ratios. ``EB Mining'' uses Equation (\ref{eq:ephat-def}) while ``Naive Mining'' uses the standard calculation. We hold for one year and repeat. 'Pub Anytime' is a portfolio that equally weights strategies from \citet{ChenZimmermann2021}. 'Pub Pre-2004' equally weighs strategies published before 2004. \textbf{Interpretation: }Rigorous data mining generates out-of-sample returns comparable to those from the best journals in finance, even if the data includes signals with zero out-of-sample mean returns, like ticker-sorted portfolios. 
\end{singlespace}

\begin{centering}
\vspace{0ex}
 
\par\end{centering}
\centering{}\setlength{\tabcolsep}{1ex} 
\centering{}{\small\begin{tabular}{lcccc}
\toprule
 & \makecell{Num Strats \\ Combined} & \makecell{Mean Return \\ (\% ann)} & $t$-stat & \makecell{Sharpe Ratio \\ (ann)} \\
\midrule
Panel A &  &  &  &  \\
\hline
EB Mining Top 1\% & 1278 & 5.70 & 9.00 & 1.46 \\
EB Mining Top 5\% & 6389 & 4.03 & 8.27 & 1.34 \\
EB Mining Top 10\% & 12777 & 2.77 & 7.16 & 1.16 \\
\hline
 &  &  &  &  \\
Panel B &  &  &  &  \\
\hline
Naive Mining Top 1\% & 1278 & 5.72 & 8.91 & 1.45 \\
Naive Mining Top 5\% & 6389 & 3.72 & 7.73 & 1.25 \\
Naive Mining Top 10\% & 12777 & 2.61 & 6.77 & 1.10 \\
\hline
 &  &  &  &  \\
Panel C &  &  &  &  \\
\hline
Pub Anytime & 203 & 5.88 & 12.54 & 2.03 \\
Pub Pre-2004 & 82 & 5.23 & 9.57 & 1.55 \\
\bottomrule
\end{tabular}
 }{\small\par}
\end{table}

\clearpage\pagebreak 
\begin{table}
\caption{\textbf{Description of the Top 1\% Data-Mined Strategies}}
\label{tab:top_strat_desc} \pdfbookmark{Table 3}{top_strat_desc}

\begin{singlespace}
\noindent Panel A shows the fraction of strategies that comes from
each signal family, pooled across all sample years. Panel B lists
the definitions of the strategies with highest predicted Sharpe Ratios
(SR pred) using data from 1974-1993. SR OOS is the realized Sharpe
ratio 1994-2003. All strategies in Panel B are equal-weighted. \textbf{Interpretation:} The top 1\% strategies are largely equal-weighted accounting strategies.
Equal-weighted past return strategies comprise a non-trivial minority.
The top 20 strategies are distant from strategies popular the academic
literature at the time of \citet{fama1993common}, yet they perform well out-of-sample. 
\end{singlespace}

\vspace{2ex}
 
\centering
\footnotesize
\setlength{\tabcolsep}{1ex}
\begin{tabular}{ccclll}
\toprule
\multicolumn{6}{c}{Panel A: Average Fraction of Signals in the Top 1\%} \\
\midrule
Acct EW & Acct VW & Past Ret EW & \multicolumn{1}{c}{Past Ret VW} & \multicolumn{1}{c}{Ticker EW} & \multicolumn{1}{c}{Ticker VW} \\
91.0\% & 0.3\% & 8.6\% & \multicolumn{1}{c}{0.1\%} & \multicolumn{1}{c}{0.0\%} & \multicolumn{1}{c}{0.0\%} \\
\midrule
\end{tabular}%
\\
\vspace{1.5ex}
\begin{tabular}{ccclll}
\midrule
\multicolumn{6}{c}{Panel B: Top 20 Strategies in 1993 based on Signed Predicted Sharpe Ratio} \\
\midrule
Rank  & \makecell{SR \\ Pred} & \makecell{SR \\ OOS} & \multicolumn{1}{c}{Signal Family} & Signal Name &  \\
\midrule
1     & 1.56  & 1.32  & Acct EW & \multicolumn{2}{l}{- $\Delta$Interest paid net / Lag(Common equity)} \\
2     & 1.51  & 0.84  & Acct EW & \multicolumn{2}{l}{- Debt due in 2nd year / Depr, depl \& amort} \\
3     & 1.43  & 0.92  & Acct EW & \multicolumn{2}{l}{- Debt mortgages \& other sec / Sales} \\
4     & 1.37  & 1.60  & Acct EW & \multicolumn{2}{l}{- Debt mortgages \& other sec / Depr, depl \& amort} \\
5     & 1.37  & 1.64  & Acct EW & \multicolumn{2}{l}{- $\Delta$Interest paid net / Lag(Stockholders equity)} \\
6     & 1.35  & 0.54  & Past Ret EW & \multicolumn{2}{l}{+ Return in quarters $t$ minus 5, 9, 17, and 18} \\
7     & 1.35  & 0.68  & Acct EW & \multicolumn{2}{l}{- Debt due in 3rd year / Depr, depl, and amort} \\
8     & 1.35  & 0.69  & Acct EW & \multicolumn{2}{l}{- Debt mortgages \& other sec / Cost of goods sold} \\
9     & 1.34  & 1.00  & Acct EW & \multicolumn{2}{l}{- $\Delta$Interest paid net / Lag(Inventories)} \\
10    & 1.33  & 0.62  & Acct EW & \multicolumn{2}{l}{- Debt mortgages \& other sec / Operating expenses } \\
11    & 1.33  & 0.47  & Past Ret EW & \multicolumn{2}{l}{+ Return in quarters $t$ minus 17} \\
12    & 1.32  & 0.62  & Past Ret EW & \multicolumn{2}{l}{+ Return in quarters $t$ minus 9, 17, 18 and 19} \\
13    & 1.30  & 0.43  & Acct EW & \multicolumn{2}{l}{- $\Delta$Liabilities / Lag(Depr \& amort)} \\
14    & 1.29  & 0.44  & Past Ret EW & \multicolumn{2}{l}{+ Return in quarters $t$ minus 9, 13, 17, and 18} \\
15    & 1.29  & 1.24  & Acct EW & \multicolumn{2}{l}{- $\Delta$Interest paid net / Lag(Equity liquidation value)} \\
16    & 1.29  & 0.68  & Past Ret EW & \multicolumn{2}{l}{+ Return in quarters $t$ minus 3, 9, 17, and 18} \\
17    & 1.25  & 0.49  & Acct EW & \multicolumn{2}{l}{- Debt due in 4th year / Depr, depl \& amort} \\
18    & 1.25  & 1.01  & Acct EW & \multicolumn{2}{l}{- Stock issuance / Gross profit } \\
19    & 1.25  & 0.55  & Acct EW & \multicolumn{2}{l}{- Debt due in 2nd year / Depr \& amort} \\
20    & 1.24  & 1.51  & Acct EW & \multicolumn{2}{l}{- $\Delta$Liabilities / Lag(Depr, depl \& amort)} \\
\bottomrule
\end{tabular}%

\end{table}

\end{document}